\def\agt{\,\raise.3ex\hbox{$>$\kern-.75em\lower1ex\hbox{$\sim$}}\,}
\def\alt{\,\raise.3ex\hbox{$<$\kern-.75em\lower1ex\hbox{$\sim$}}\,}

\documentstyle[epsfig]{mnkop}

\title{Halo structure,  masses of dark objects and parallax microlensing} 
\author[D. Markovi\'{c}]{D. Markovi\'{c} \\
    Theoretical Astrophysics Center, Juliane Maries Vej 30, DK-2100
    Copenhagen \O, Denmark~\footnote{}\\
        draza@tac.dk}
\begin{document}
\input{epsf}

\maketitle

\begin{abstract}
We study the use of parallax microlensing to separate the
effects of the mass function of dark massive 
halo objects (MHOs or `machos') on 
the one hand and their spatial distribution 
and kinematics on the other.  This disentanglement is supposed
to allow a much better determination of the two than could be achieved
entirely on the basis of the durations of events.  
We restrict our treatment to the same
class of power-law spherical models for the halo of 
MHOs studied in a previous paper~\cite{paper.I}.  
Whereas
the duration-based error in the average MHO mass, 
$\bar{\mu}\equiv\bar{M}/M_{\odot}$ exceeds (at $N=100$ events) 
$\bar{\mu}$ by  a factor of 2
or more, parallax microlensing remarkably 
brings it down to 15-20\% of $\bar{\mu}$,
regardless of the shape of the mass function.  In addition, the slope 
$\alpha$ of the mass function, $dn/d\mu\propto\mu^{\alpha}$,
can be inferred relatively accurately ($\sigma_{\alpha} < 0.4$) for
a broader range, $-3 <\alpha < 0$.
The improvement in the inference of the halo structure is
also significant: the index $\gamma$ of the density profile 
($\rho\sim R^{-\gamma}$)
can be obtained with the error $\sigma_{\gamma} <0.4$.  While
in a typical situation
the errors for the parameters specifying the velocity dispersion
profile are of about the same magnitude as the parameters, 
virtually all the uncertainty is `concentrated' in linear
combinations of the parameters that may have little
influence on the profile and thus allow its reasonably
accurate inference.
\end{abstract}

\begin{keywords}
Microlensing -- Galactic halo -- Macho mass function.
\end{keywords}

\section{Introduction and overview}
\footnotetext {Address after  September 1st, 1997:  
        Department of Physics,
	 University of Illinois at Urbana-Champaign, 
	 1110 W. Green St., Urbana, IL 61801, USA}
A statistical analysis by Alcock {\it et al.} (1996) of the 2-year
microlensing data (6 or 8 events) obtained by the MACHO project
in the direction of the
Large Magellanic Cloud (LMC)  indicated that the massive dark halo
objects (MHOs or `machos') responsible for the microlensing events could account 
for 30-100\% of the total mass in the halo of our Galaxy.  
According to their analysis,
typical (average) mass of the MHOs should lie in the range 
$0.1-0.6\, M_{\odot}$.
The more recent 4-year data (14 events, Axelrod 1997) yield similar ranges
of the inferred quantities.  
Apart from the statistical error due to the relatively small
number of events, our ignorance regarding the structure of the halo
of massive objects (i.e., their spatial distribution and kinematics)
leads to rather large uncertainties in the inferred masses.

This last source of error is not likely to be extinguished if
one relies only on the measurement of event durations $T=R_{\rm E}/v_{\rm n}$
($R_{\rm E}$ is the Einstein radius, $v_{\rm n}$ is the MHO's velocity
orthogonal to the line of sight).  Indeed, as shown
by Markovic \& Sommer-Larsen (1997, paper I), for the number of
events $N < 1000$ the halo structure cannot be constrained sufficiently
to allow a determination of the average mass 
$\bar{\mu}\equiv \bar{M}/M_{\odot}$  of the MHOs to better
than a factor of about 2.   
  Furthermore, paper I discussed only a limited class of
spherical haloes; the results of Evans (1997), based on a far wider
variety of halo models, imply that the range of $\bar{\mu}$ (at virtually
arbitrary $N$) could in principle extend from 0.1 to 1.

The duration $T$ is, however, not the only relevant quantity that
can be obtained from a microlensing event.  For instance, photometric
(Gould 1994a; Nemiroff \& Wickramasinghe 1994) or spectroscopic
(Maoz \& Gould 1994) methods have been proposed to measure 
the proper motion of the lens ${\bf v}_{\rm n}/zD$, where
$D$ is the Earth-source distance, $z\equiv D_{\rm L}/D$ and
$D_{\rm L}$ is the Earth-lense distance. 
Another approach [discussed by  
Grieger, Kayser \& Refsdal (1986) in the context of quasar astronomy] 
is parallax microlensing, i.e., observing magnification
through telescopes displaced from each other by about 1AU.  
More recently, Gould (1994b) studied and advocated the
use of parallax microlensing to obtain more information regarding
the position and velocities of the lenses (and consequently reduce
the uncertainty of their masses).  

The utility of the parallaxes stems primarily from the fact
that the delay $\tau$ between the maximal magnifications 
in the two detectors (one on the Earth and the other
on a satellite in a heliocentric orbit) does not depend on the mass
of the MHO crossing the two lines of sight to a source 
Additional information
is contained in the two maximal magnifications determined
by the impact parameters $u_1$ and $u_2$ measured in units
of the Einstein radius.  Breaking a 4-fold degeneracy (Gould 1994b;
see also Section 2 of the present paper) by observing from
a second satellite would allow us to obtain the so-called 
reduced transverse velocity $\tilde{\bf v}\equiv {\bf v}_{\rm n}/(1-z)$
of the MHO.  A sufficiently large number of such measurements 
would then presumably put tight constraints on the structure of the halo.  
However, even in absence of a second satellite, the relative motion
of the first satellite and the Earth could suffice to reduce
the ambiguity to (at most) a 2-fold degeneracy
in the direction of $\tilde{\bf v}$ for a
majority of events (Gould 1995; Boutreux \& Gould 1996).  [According
to Han \& Gould (1995) this should (in the case of galactic
 bulge microlensing) permit measurement of {\it individual} MHO 
 masses to an accuracy of about  0.2 on the logarithmic scale.]

In this paper we explore quantitatively the extent to which one
could expect parallax microlensing to help
constrain the mass function of the MHOs as well as  their
spatial distribution and kinematics.
The assumptions of the present paper are similar to those 
of paper I: for convenience we again limit ourselves to a class of
spherical halo models (see section 3) described by a set of 5
parameters (the singular isothermal sphere is a particularly
simple member of this class).  On the other hand, the mass function
is assumed to be a simple power law, $d n/d\mu \propto \mu^{\alpha}$
characterised by three parameters (independent of
the position in the halo): the average mass $\mu$, slope $\alpha$
and range $\beta$ on the logarithmic scale.

Although lacking somewhat in generality, this framework will allow
a straightforward application of the apparatus of statistical
parameter estimation: the errors of maximum likelihood inference
of the mass function and halo parameters can be estimated from the
sensitivity of the distributions of directly measurable quantities
to small shifts in the underlying parameters (see paper I and
Section 5 of the present paper).  We call such estimates the
Cramer errors \cite{cramer}.  For simplicity we will study both the 
{\it degenerate} (with the full 4-fold degeneracy) parallax
microlensing, where 
the observable quantities are $T$, $p\equiv\tau/T$, $u_1$ and $u_2$
and {\it resolved} (the 4-fold degeneracy completely removed)
parallax microlensing with observables $T$, $p$ and $w=\sqrt{a^2 - p^2}$,
where $a$ is the transversal distance between the two lines of sight
in the lens plane measured in Einstein radii.

In specific computations we at first adopt
for the parameters of the underlying halo model 
the values corresponding to the centrally 
condensed ($\gamma =3.4$; $\rho\propto R^{-\gamma}$; $R$ is
the distance from the centre of the Galaxy) halo of 
blue horizontal branch field stars (BHBFS; see Section 3).
Although one might speculate as to the relevance 
--- or irrelevance --- of this 
structure to the halo of MHOs (paper I), these values
are simply taken as a convenient starting point for our numerical
experiment and the accuracy of their inference is estimated.
In addition we briefly discuss the inference starting
from the singular isothermal sphere ($\gamma =2$; constant,
isotropic velocity dispersion) as a model for the MHO
halo.

We find that parallax microlensing reduces the Cramer errors 
in $\bar{\mu}$ from a factor of 2-10, characteristic of measurements
of event durations, to only 15-20\% (at $N=100$ events)
for {\it a priori} unknown halo structure parameters.  
This error is typical
of inference under the (unrealistic) assumption that the halo model
{\it is} accurately known {\it a priori}
and is kept fixed in the maximum-likelihood
fitting of the distribution of measurable quantities.  The improvement
indeed results from an effective disentanglement of the mass function
from the halo structure.  In addition,
while parallax-based errors in parameters $\alpha$ and $\beta$,
specifying the shape of the mass function, are comparable to
duration-based errors if $\alpha$ is sufficiently close to -1.5,
the growth of the errors away from this value is strongly restrained by 
the parallaxes (see Fig.~10).

Parallax microlensing also reduces by about two orders of magnitude
the Cramer errors for the halo parameters.  The power index $\gamma$ of
the halo density profile is determined with
the error $\sigma_{\gamma}< 0.4$ (at both $\gamma =3.4$ and $\gamma =2$), 
again from $N=100$ events.  
On the other hand,
the errors in the parameters specifying the velocity dispersion profile
are of roughly the same magnitude as the parameters themselves.
This, however, does not necessarily mean that the profile will be poorly
determined: as shown in Section 5, virtually all uncertainty 
(at $\gamma =3.4$)
is due to the existence of a {\it single} 
linear combination of velocity dispersion 
parameters that is poorly constrained even by parallax microlensing.
Indeed, small displacements in the parameter space along this vector cause 
a particularly  weak change in the velocity profile.  
In addition, these changes
tend to occur predominantly at large radii,
where the microlensing rate is low.  The peculiarities of
the singular isothermal sphere, on the other hand, lead
to the existence of {\it three} poorly constrained linear combinations
of velocity dispersion parameters, none of them
having a significant effect on the flat velocity profile.  Remarkably,
all the above conclusions are virtually independent of whether we use the
degenerate  or fully resolved  parallaxes:  the improvement due
to resolving the degeneracy is modest. 

In Section 2 of this paper we derive expressions for distribution
functions of measurable quantities.  These expressions are general 
and can be used with arbitrary halo models.
The specific class of halo models used in this paper is
described in Section 3 (following a similar section in paper I).
Section 4 deals with the morphology of the distribution functions
derived in Section 2, while Section 5 explores their sensitivity to the
underlying parameters and thus derives the Cramer errors of inference.
Finally, section 6 contains basic conclusions of this
paper along with a bit of speculation regarding their more 
general validity.

\section{Differential parallax microlensing rates}

As observed from the Earth, a lens of mass $M=\mu M_{\odot}$, crossing
the Earth-source line of sight at distance $z D$ ($0 \leq z\leq 1$) 
from the Earth and with the impact parameter $u_{1} R_{\rm E}$  [$R_{\rm E} =
r_{\rm E}\sqrt{\mu}\sqrt{z(1-z)}$, $r_{\rm E} \equiv 2\sqrt{G M_{\odot}D/c^2} = 
3.2\times 10^{9}\,$km] will magnify the star by the maximum factor
$A_{\rm max} = (u_{1}^{2} + 2)/(u_1 \sqrt{u_{1}^2 +4}\,)$.  
On the other hand, a satellite will
detect maximum magnification, determined by the impact parameter 
$u_2 R_{\rm E}$ relative to the satellite-source line of
sight, with the time shift $\tau$ from the moment of the Earth-observed
maximum magnification.

\begin{figure}[t]
\label{PWgeom}
\centering
\hbox{\epsfxsize= 8 cm \epsfbox[40 150 570 570]{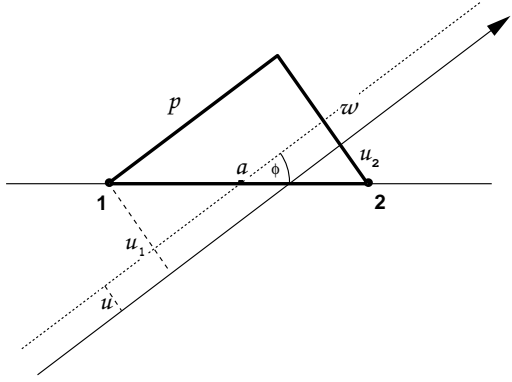}}
\caption{A parallax microlensing event projected onto the lens plane  
(orthogonal to the line of sight). All distances are given in units of the
Einstein radius $R_{\rm E}$.}
\end{figure}

The geometric relations between the observable quantities, 
$T\equiv R_{\rm E}/v_{\rm n}$ (event duration as observed by both
the Earth-based observer and the satellite), $\tau$, $u_1$ and $u_2$,
are easily derived from Fig.~1.
If $\bf{r}$ is the component of the Earth-satellite vector orthogonal
to the line of sight, its source-centered projection onto the 
plane (also orthogonal to the line of sight) of the lens is
\begin{equation}
{\bf a} = \frac{{\bf r} (1-z)}{R_{\rm E}} =
  \frac{\bf r}{r_{\rm E}}\frac{1}{\sqrt{\mu}}\sqrt{\frac{1-z}{z}}
\end{equation}
(measured in units of the Einstein radius $R_{\rm E}$). 

\begin{figure}[t]
\label{PWdeg}
\centering
\hbox{\epsfxsize= 8 cm \epsfbox[-10 120 550 670]{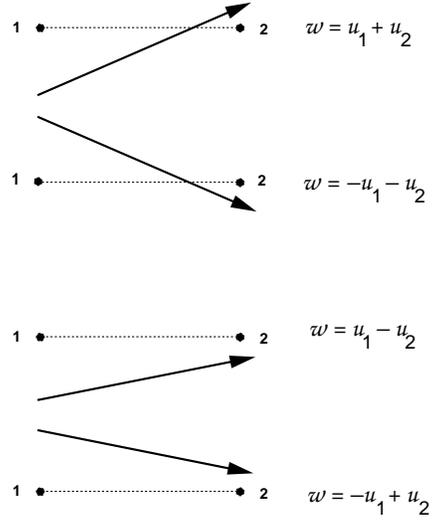}}
\caption{Four possibilities corresponding to a pair of values $u_1 \geq 0$,
 $u_2 \geq 0$ in the case of {\it degenerate} parallax microlensing. }
\end{figure}

The points `1'  and `2' in Fig.~1 denote intersections
of the Earth-source and satellite-source lines of sight respectively
with the lens plane.
The lens' trajectory along the unit vector $\hat{\bf v} = 
(\cos\phi, \sin\phi)$ (again, projected on the lens plane), shown as
the solid arrow, crosses the 1-2 line at distance $u R_{\rm E}$
from the parallel line (dotted) drawn through the midpoint between
1 and 2.  Consequently, 
\begin{equation}
\label{p.def}
p\equiv a\cos\phi = \frac{r(1-z)\cos\phi}{v_{\rm n}} \frac{1}{T} =
\frac{\tau}{T}
\end{equation}
is an observable quantity.  On the other hand, $w\equiv a\sin\phi$
generally cannot be obtained unambiguously from $u_1$ and $u_2$ only;
the 4-fold degeneracy is illustrated in Fig.~\ref{PWdeg}.
A secure way of breaking this degeneracy would be to use a second satellite
[the line tangent to three circles of radii $u_1$, $u_2$ and $u_3$ (the
last measured from the second satellite)
is unique].  However, as Gould (1995) has shown,
the motion of the Earth and the single satellite relative to
the line of sight should allow us to resolve the ambiguity --- at least
regarding the magnitude of $w$ --- in most cases.  Nevertheless, in this 
paper we will discuss both the `{\it degenerate}' (the 4-fold ambiguity 
unresolved) and the `{\it resolved}' ($w$ uniquely determined) 
parallax microlensing.

The lens' rate of crossing near the line of sight per single source 
and a single (number density near the Sun $n_o =1$) lens is 
\begin{eqnarray}
\label{total.rate}
\Gamma =\int d\mu \frac{dn_o}{d\mu}\int D H(z) dz\int R_{\rm E}(z) du
\int f_{\rm n}(v_{\rm n},\phi ) v_{\rm n}^2 d v_{\rm n} d\phi,\hspace{-5cm}
 \nonumber \\ 
\end{eqnarray}
where $H(z)$ [$H(0)=1$] is the MHOs' halo density profile along
the line of sight, $f_{\rm n}$ 
[$\int f_{\rm n}(v_{\rm n},\phi ) v_{\rm n} d v_{\rm n} d\phi = 1$]
is the $z$-dependent, 2-dimensional
distribution of velocities projected orthogonal to the line of sight and 
$dn_o /d\mu$
[$\int (dn_o /d\mu)d\mu = 1$] is the $z$-independent mass function of
the MHOs.

Using
\begin{equation}
\label{z.a}
z = \frac{(r/r_{\rm E})^2}{\mu a^2 + (r/r_{\rm E})^2},
\end{equation}
rewriting the lens plane area element
\begin{equation}
\label{dPdW}
dp dw = a da d\phi = \frac{1}{2} \left(\frac{dz}{da^2}\right)^{-1} dz d\phi,
\end{equation}
and switching to integration over $T$ ($d v_{\rm n}/dT = - v_{\rm n}/T$)
we obtain
\begin{eqnarray}
\label{chi}
& & \hspace{-0.6cm}\chi (T,p,w) \equiv \frac{d\Gamma}{dT dp dw} \nonumber \\
& &  \nonumber \\
& & \hspace{-0.5cm}= \int d\mu \frac{dn_o}{d\mu} DH(z) 
  R_{\rm E}(z) f_{\rm n}(v_{\rm n},\phi )
 \frac{v_{\rm n}^3}{T}  
 \frac{2 (r/r_{\rm E})^2 \mu}{\left[\mu a^2 + (r/r_{\rm E})^2\right]^2},
  \nonumber \\
\end{eqnarray}
where $v_{\rm n} = r_{\rm E}\sqrt{\mu}\sqrt{z(1-z)}/T$ and $\tan\phi = w/p$.

By contrast with resolved parallaxes, where $\chi(T,p,w)$ is
of more immediate relevance, measuring $u_1 = |u +w/2|$ and $|u -w/2|$
(see Fig.~1) is not sufficient to uniquely determine $w$
in the case of degenerate parallaxes. Inserting `dummy' integration
$\int du_1 \delta (u_1 - |u +w/2|) \int du_2 \delta (u_2- |u -w/2|)$
in the rate (\ref{total.rate}) and using the identity
$ d(u + w/2)\wedge d(u-w/2) = dw\wedge du$,
we arrive at the expected result
\begin{eqnarray}
\label{degen.rate}
\Psi (T,p,u_1 ,u_2) &\equiv& \frac{d\Gamma}{dT dp du_1 du_2} \nonumber \\
	\nonumber \\
& &\hspace{-2.2cm} =\; \chi (T,p, u_1 + u_2) + \chi (T,p, -u_1 - u_2) 
   \nonumber \\
& &\hspace{-2cm} + \;\chi (T,p, u_1 - u_2) + \chi (T,p, -u_1 + u_2),
\end{eqnarray}
expressing the differential rate in terms of variables  accessible to 
degenerate parallax microlensing detection.

So far in this section we have ignored the question of the minimum
amplification necessary for successful detection of a microlensing 
event.  In particular, microlensing might produce
sufficient magnification only in the Earth- (or satellite-) based
detector, i.e., the magnification in the other detector (say `2') 
could be too small for a reliable determination of $\tau$ and $u_2$. To
deal with this possibility, we will require $u_1 < u_{\rm th}$, 
$u_2 < u_{\rm th}$ for a detectable parallax (double, i.e. in
both detectors) microlensing event, while single events will 
correspond to $u_1 < u_{\rm th}$, $u_2 > u_{\rm th}$ or 
$u_1 > u_{\rm th}$, $u_2 < u_{\rm th}$, where $u_{\rm th}$
is a certain threshold value.  The {\it detection} rate of events 
(both single and double) in one (say `1') detector is thus
\begin{eqnarray}
\label{one.rate}
P(T, u_1) &&\hspace{-0.5cm}\equiv \frac{d\Gamma^{(1)}}{dT du_1} \nonumber \\
 &&\hspace{-0.5cm}=\int_{0}^{\infty} du_2 
 	\int_{-\infty}^{\infty} dp\,\Psi(T,p,u_1, u_2)
 	\nonumber \\
 &&\hspace{-0.5cm}=\int_{0}^{u_{\rm th}} du_2 
 	\int_{-\infty}^{\infty} dp\,\Psi(T,p,u_1, u_2)
   + \Omega^{(1)}(T, u_1), \hspace{-2cm} \nonumber \\
\end{eqnarray}
where
\begin{eqnarray}
\label{single.rate}
\Omega^{(1)}(T,u_1)\equiv\frac{d\Gamma^{(1)}_{\rm single}}{dT du_1}
   = \int^{\infty}_{u_{\rm th}}du_2 
   \int_{-\infty}^{\infty} dp\,\Psi(T,p,u_1, u_2) \hspace{-2cm}
    \nonumber \\
\end{eqnarray}
is the differential detection rate for single events, expressed in
terms of the only available measurables, $T$ and $u_1$.
Of course, $\Omega^{(1)}(T,u) = \Omega^{(2)}(T,u) = 
\Omega(T,u)$.
The differential rate $P(T,u)$, introduced in equation~(\ref{one.rate}),
is
\begin{eqnarray}
\label{one.rate.1}
P(T, u) &&\hspace{-0.5cm} =
2D r_{\rm E}^4 \int d\mu \frac{dn_o}{d\mu} \left(\frac{\mu}{T^2}\right)^2
  \int_{0}^{1} dz H(z) [z(1-z)]^2 \nonumber \\
  && \times \int_{0}^{2\pi}d\phi f_{\rm n}(v_{\rm n},\phi).
\end{eqnarray}

If we assume (as we will in the present paper) that the MHO mass 
function can be well approximated by a simple power law
\begin{equation}
\label{power.law}
\frac{dn_o}{d\mu} =\frac{1}{C_{\beta}(\alpha)}
   \frac{\mu^{\alpha}}{\mu_{o}^{\alpha +1}},
\end{equation}
where $\beta =\log_{10}(\mu_{\rm max}/\mu_{\rm min})$, $\mu_{\rm max}$
and $\mu_{\rm min}$ are the upper and lower bounds of the mass range,
$\mu_o = \sqrt{\mu_{\rm max}\mu_{\rm min}}$ and
\begin{eqnarray}
\label{cbetalp}
C_{\beta}(\alpha) = \left\{\begin{array}{ll}
            \beta\ln 10 & \mbox{$\hspace{0.5cm}\alpha =-1$},\hspace{-1cm}
            \nonumber \\
            \nonumber \\
            \frac{1}{\alpha +1}\left[10^{\beta(\alpha +1)/2}
              - 10^{-\beta(\alpha +1)/2}\right]
              & \mbox{$\hspace{0.5cm}\alpha \neq -1$},\hspace{-1cm}
                             \nonumber
                 \end{array}
                 \right. \nonumber \\
\end{eqnarray}
the one-detector rate $P(T,u)$ simplifies to
\begin{equation}
\label{power.law.rate}
P(T,u) = 2D r_{\rm E}^4 \frac{T^{2(\alpha +1)}}{C_{\beta}(\alpha)
  \mu_{o}^{\alpha +1}}
  \int_{\mu_o \frac{10^{-\beta/2}}{T^2}}^
          {\mu_o \frac{10^{\beta/2}}{T^2}} y^{\alpha} F(y) dy,
\end{equation}
where $y\equiv\mu/T^2$ and
\begin{eqnarray}
\label{Fy}
F(y) &=& y^2 \int_0^1 dz\; [z(1-z)]^2
  H(z) \nonumber \\
  && \hspace{1.8cm} \times\;
     \int^{2\pi}_{0} d\phi\;
     f_{\rm n} \left[r_{\rm E}\sqrt{z(1-z) y},\,\phi\right].
   \nonumber \\
\end{eqnarray}

In the rest of the paper we will use a `composite' notion of
microlensing event including double (parallax) events and
single events detected only in detector `1' or `2'.  The
composite probability distribution function for
degenerate parallax microlensing can then be
obtained by introducing the normalising constant $A$
\begin{eqnarray}
\label{hat.psi}
\hat{\Psi}(T,p,u_1,u_2) &=& \frac{1}{A} \Psi (T,p,u_1,u_2),
	\nonumber \\
\hat{\Omega}(T,u) &=& \frac{1}{A}\Omega(T,u),
\end{eqnarray}
so that
\begin{eqnarray}
\label{degen.norm}
\int^{\infty}_{0} dT \int^{\infty}_{-\infty} dp 
  \int^{u_{\rm th}}_{0} du_1 \int^{u_{\rm th}}_{0} du_2\,
  \hat{\Psi}(T,p,u_1,u_2) \nonumber \\
  + \; 2 \int^{\infty}_{0} dT \int^{u_{\rm th}}_{0} du\, 
  \hat{\Omega}(T,u) = 1.
\end{eqnarray}

In order to take account of the detection condition $u_1 < u_{\rm th}$,
$u_2 < u_{\rm th}$ in the case of {\it resolved} parallax microlensing
detection, we multiply $\chi(T,p,w)$ by the range of $u$ (see
Fig.~1) for which the double event detection condition
is satisfied
\begin{equation}
\label{resolv.rate}
\chi(T,p,w) \longrightarrow (2u_{\rm th} - |w|)\chi(T,p,w),
\end{equation}
($|w| < 2u_{\rm th}$), and thus obtain the differential
{\it detection} rate of events characterised by the observables 
$T,p$ and $w$.
The relevant probability distribution for the detectable double events
is then
\begin{equation}
\label{resolv.rate.1}
\hat{\chi}(T,p,w)= \frac{1}{A}(2u_{\rm th} -|w|)\chi(T,p,w),
\end{equation}
where $A$ has the same value as in the degenerate parallax case
[one can show  $\int^{u_{\rm th}}_{0} du_1 \int^{u_{\rm th}}_{0} du_2\,
  \Psi(T,p,u_1,u_2) = \int^{2 u_{\rm th}}_{-2 u_{\rm th}} dw\,
   (2u_{\rm th} -|w|)\chi(T,p,w)$],
thus yielding the normalisation
\begin{eqnarray}
\label{resolv.norm}
&&\hspace{-0.6cm}\int^{\infty}_{0} dT \int^{\infty}_{-\infty} dp 
  \int^{2u_{\rm th}}_{-2u_{\rm th}} dw\,
  \hat{\chi}(T,p,w) \nonumber \\  
&&  \hspace{1cm}+ \; 2 \int^{\infty}_{0} dT 
           \int^{u_{\rm th}}_{0} du\, \hat{\Omega}(T,u) = 1.
\end{eqnarray}

\section{Models of MHO distribution and kinematics}

\begin{figure}
\label{vel.disp}
\centering
\centerline{\epsfxsize= 8 cm \epsfbox[290 30 600 430]
      {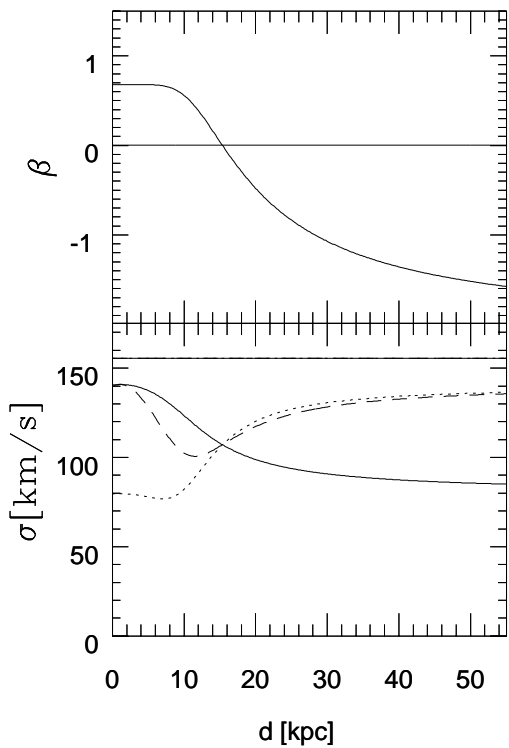}}
\caption{ Anisotropy parameter $\beta$ and velocity dispersion for
 the CS halo model as functions of the distance $d$ from the Earth in the
 direction of LMC; $\sigma_{\rm r}$ is given by the solid 
 line, $\sigma_{\rm t}=\sigma_j$
 by the dotted line and $\sigma_i$ [see paragraph following
 equation\ (\ref{fn})]
 by the dashed line.  The straight solid lines correspond to the
 singular isothermal sphere, SIS ($\beta = 0$, $\sigma = 156$ km/s) .
 }
\end{figure}

\begin{figure}[t]
\label{dPdT.ps}
\centering
\hbox{\epsfxsize= 10 cm \epsfbox[30 130 710 640]{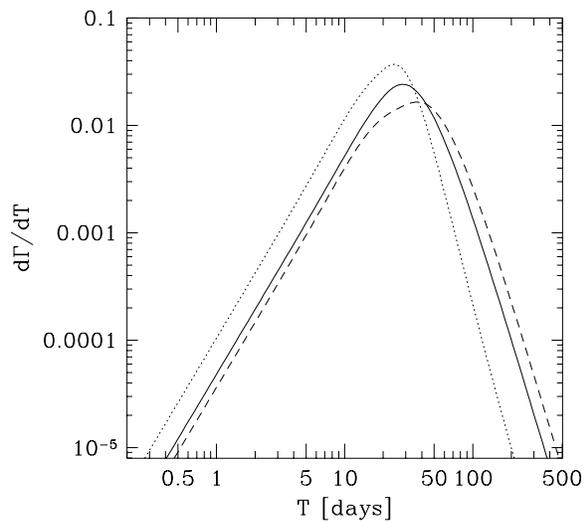}}               
\caption{Differential rate $d\Gamma/dT$ (normalised) for the
SIS (solid line) and CS (short-dashed) halo models 
with Sun's and LMC motion neglected.  
For the dotted line one assumes the CS model and takes into account the
motion of the Sun and LMC.
In all cases
the mass function is a delta function centered on $\mu =0.4$.
}
\end{figure}

In this paper we will consider a range of spherically symmetric
models  of the massive halo
objects' distribution and velocities.  
Probably the most commonly used
model is the isothermal sphere with the
velocity dispersion constant throughout the halo and the density profile
which is well approximated by
\begin{equation}
\label{isoth.core}
\rho (R) = \rho_o\frac{a^2 + R_\odot^2}{a^2 + R^2},
\end{equation}
where $a\approx 5\,$kpc is the `core' radius and $R_\odot = 8.5\,$kpc is
the distance of the Sun from the galactic centre.   Assuming that the total
 (luminous
+ dark matter) halo density is distributed according to expression
(\ref{isoth.core}), one obtains the observed (approximately) flat rotation
curve for the galaxy.

The MHO mass distribution, however, need not follow that
of the total halo mass. 
We may, for instance, follow the hints provided by
recent observations (Sommer-Larsen, Flynn \& Christensen 1994,
Sommer-Larsen {\it et al.} 1997) of the blue horizontal
branch field stars (BHBFS) in the outer halo.  These observations
imply that the velocity dispersion changes from $\beta \equiv
1 - \sigma_{\rm t}^2 /\sigma_{\rm r}^2 > 0$ ($\sigma_{\rm r}$ and
$\sigma_{\rm t}$ are velocity dispersions respectively in
the radial and tangential direction relative to the Galactic centre)
at smaller distances $R$
from the centre of the Galaxy to $\beta < 0$ at larger distances.
The radial velocity dispersion is well described
by the analytic fit
\begin{equation}
\label{disp.fit}
\sigma_{\rm r}^2 = \sigma_o^2 + \sigma_{+}^2 \left[\frac{1}{2} -
 \frac{1}{\pi}\tan^{-1}
 \left(\frac{R-r_o}{l}\right)\right],
\end{equation}
where the best agreement with the observations is achieved with
$\sigma_o = 80\, {\rm km}\,{\rm s}^{-1}$, $\sigma_{+} = 145\,
 {\rm km}\,{\rm s}^{-1}$, $r_o = 10.5\,$kpc and $l = 5.5\,$kpc
(these are the values used in paper I and the present one; 
more recent values, based on
 a larger sample of stars, are given in Sommer-Larsen {\it et al.} 1997).
The BHBFS halo is close to spherical with the
density that is well modeled by the power law $\rho = \rho_o
(R_{\odot}/R)^{\gamma}$, where $\gamma\approx 3.4$.

\begin{figure*}[t]
\label{distr.SIS}
\centering
\hbox{\epsfxsize= 6.5 cm \epsfbox[70 250 350 720]{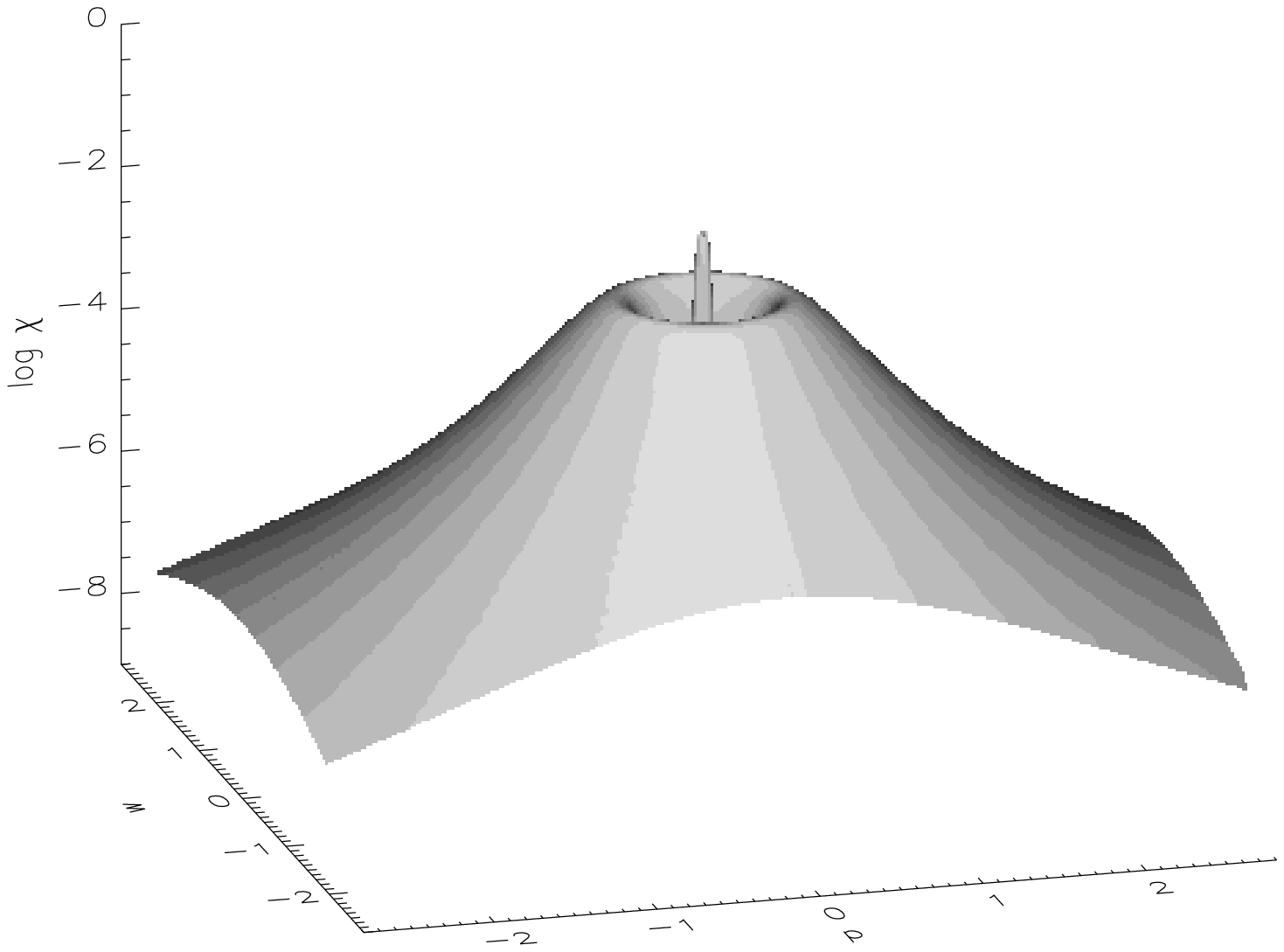}
      \epsfxsize= 5 cm \epsfbox[-280 -170 150 570]{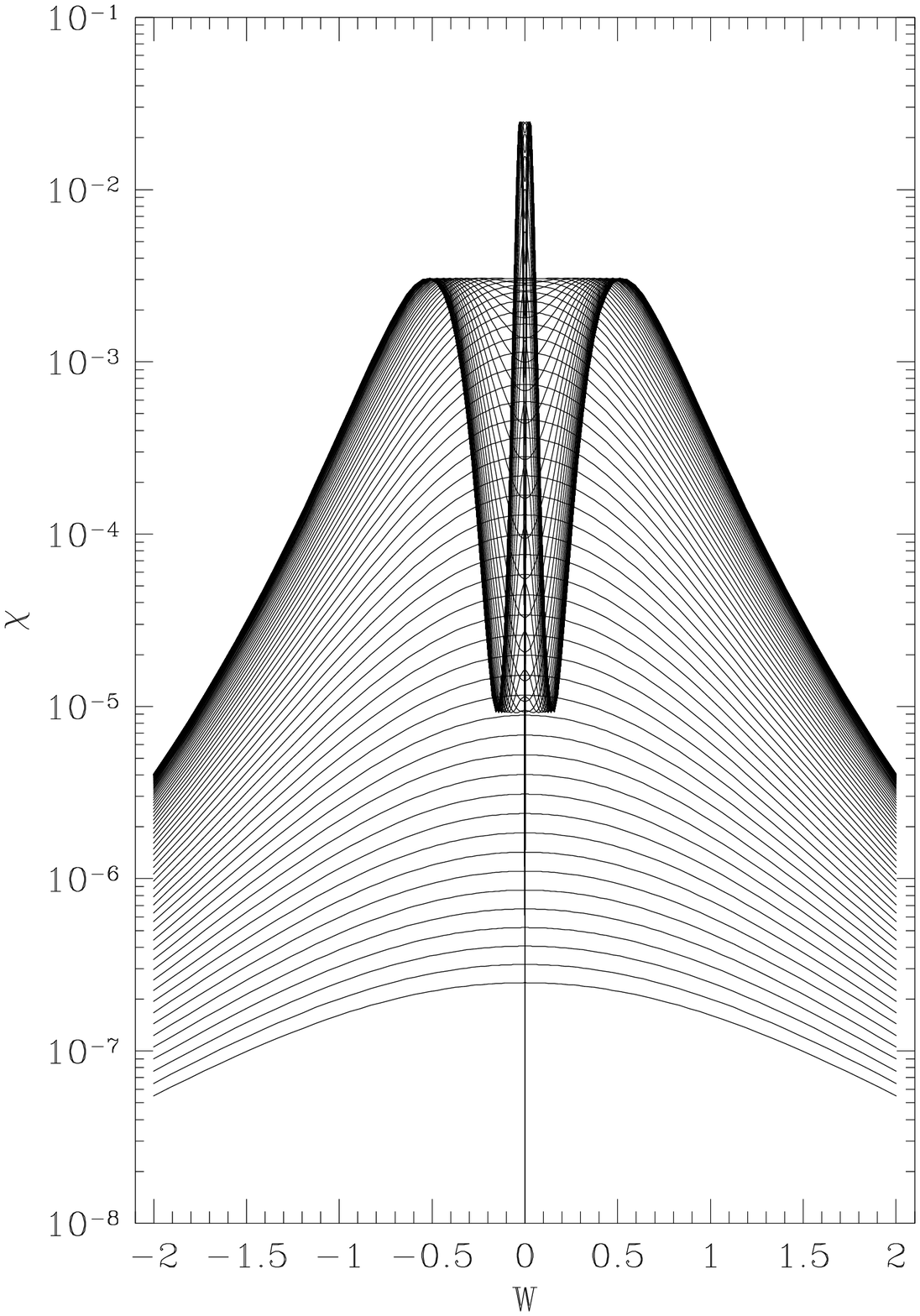}}
\hbox{\epsfxsize= 6.5 cm \epsfbox[70 350 350 600]{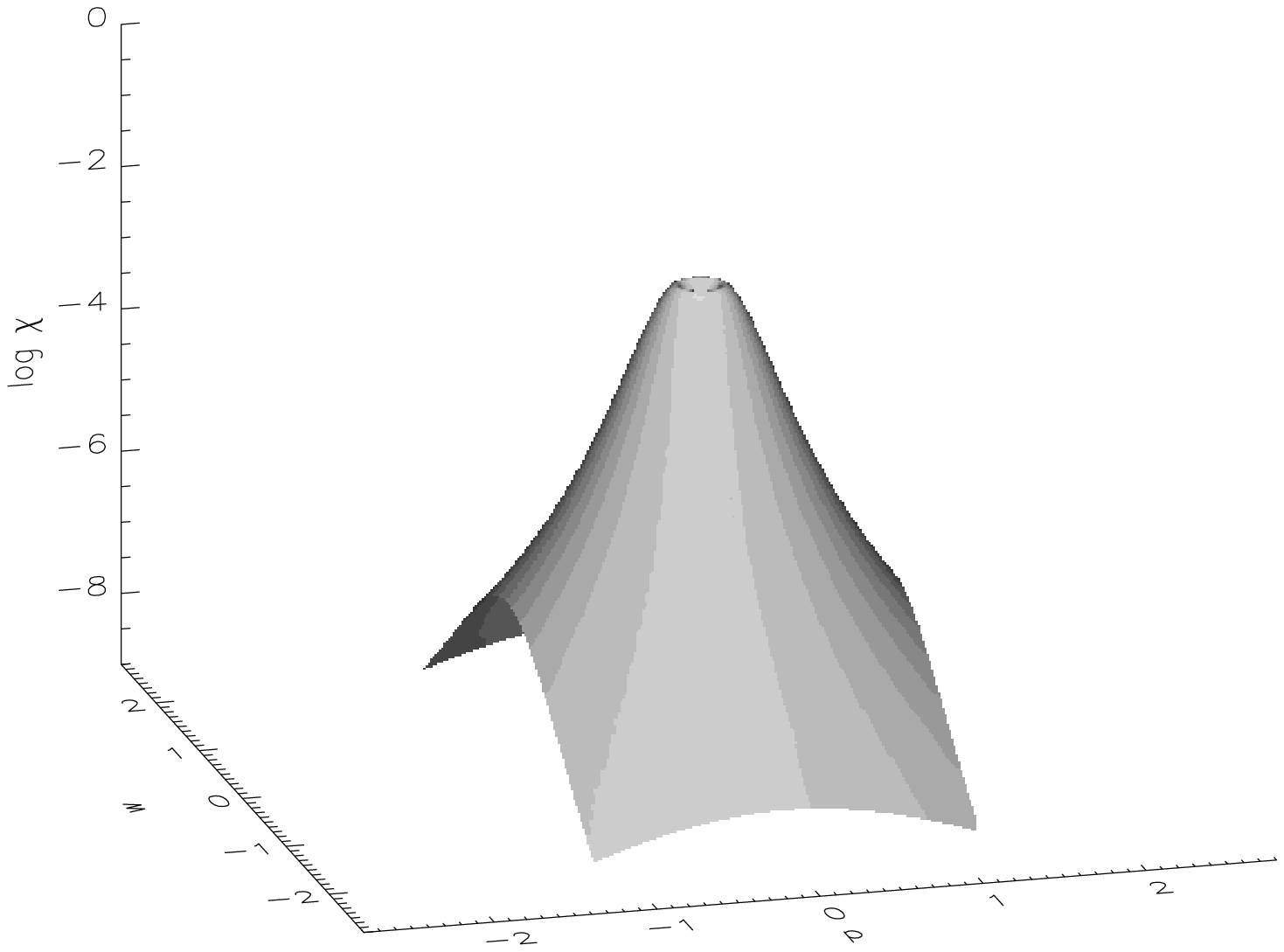}
      \epsfxsize= 5 cm \epsfbox[-280 30 150 220]{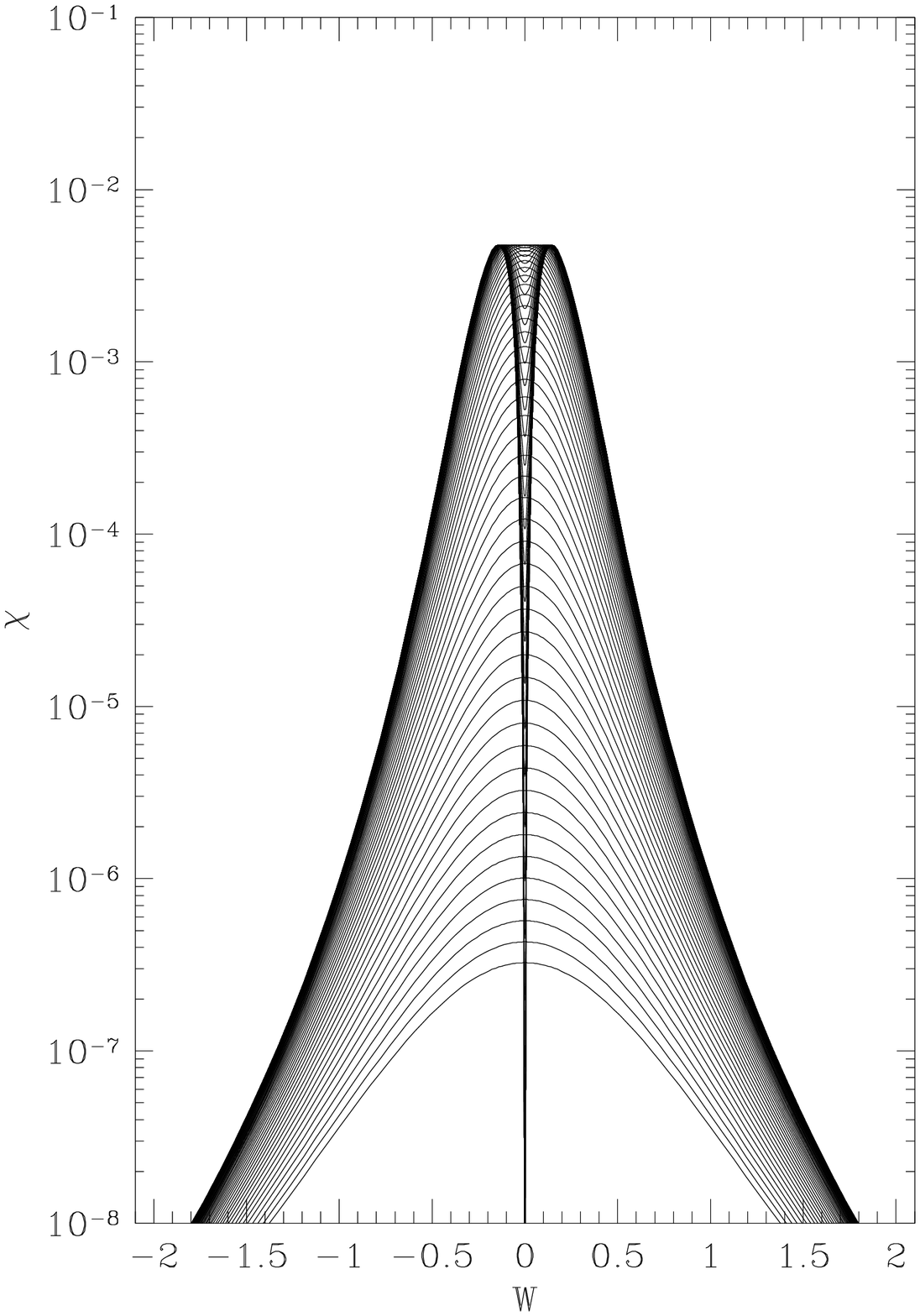}}           
\caption{Plots of $\chi$ with the SIS halo (Sun's and LMC motion
neglected) and a delta mass function
($\mu = 0.4$) at $T=14.2$ (top) and
$T=84.7$ (bottom) days.
}
\end{figure*}

The Jeans' equation for spherical systems \cite{binney}
yields the tangential velocity dispersion
\begin{equation}
\label{disp.tan}
\sigma_{\rm t}^2 = \frac{1}{2}V_{\rm c}^2 - \left(\frac{\gamma}{2} -
   1\right)\sigma_{\rm r}^2 + \frac{r}{2}\frac{d\sigma_{\rm r}^2}{dR},
\end{equation}
where $V_{\rm c} = (-R d\Phi/dR)^{1/2}$ is the (roughly constant) rotation
velocity.
This tangential dispersion is
smaller than in the case of an isothermal sphere ($\gamma = 2$,
$\sigma_{\rm r} =$ const.) with the same $V_{\rm c}$
(see Fig.~\ref{vel.disp}).

We will, following paper I, model the
velocity distribution by the Gaussian
\begin{eqnarray}
\label{gaussian}
f(v_r, v_{\theta}, v_{\phi}) = \frac{1}{(2\pi)^{3/2}}
   \frac{1}{\sigma_{\rm r}\sigma_{\rm t}^2}\exp\left[-\frac{1}{2}\left(
   \frac{v_r^2}{\sigma_{\rm r}^2} +
   \frac{v_{\theta}^2 + v_{\phi}^2}{\sigma_{\rm t}^2}\right)\right],
    \hspace{-4cm} \nonumber \\
\end{eqnarray}
where $\sigma_{\rm r}$ and $\sigma_{\rm t}$ are given by equations\
(\ref{disp.fit}) and\ (\ref{disp.tan}) for
power-law density profiles.  The relevant distribution [used, e.g., 
in the rate~(\ref{total.rate})]
of velocities orthogonal to the line of sight is then
\begin{eqnarray}
\label{fn}
f_{\rm n}(v_{\rm n},\phi) = \frac{1}{2\pi\sigma_i\sigma_j}
  \exp\bigg[-\frac{1}{2}\bigg(& &\hspace{-0.7cm}
    \frac{[v_{\rm n}\cos(\phi +\phi_o) + s_i]^2}
  {\sigma_{i}^2} 
  \nonumber \\
  & &\hspace{-0.7cm}+ \frac{[v_{\rm n}\sin(\phi +\phi_o) + s_j]^2}
  {\sigma_{j}^2}\bigg)\bigg]. \hspace{-1cm}
   \nonumber \\
\end{eqnarray}
For the above equation we have introduced orthonormal vectors in
the plane orthogonal to the line of sight: $\hat{\bf\imath}$
is in the plane determined by the Sun, LMC and the Galactic centre
(GC) and points in the general direction of GC; $\hat{\bf\jmath} =
\hat{\bf k}\times\hat{\bf\imath}$, where $\hat{\bf k}$ points
along the Sun-LMC line of sight.  Thus, $s_i$ and $s_j$ are
the corresponding components of the local ($z$-dependent) velocity
of the line of sight relative to the galaxy, $\phi_o$
is the angle between the `1'-`2' axis and $\hat{\bf\imath}$,
$\sigma_j = \sigma_{\rm t}$ and 
$\sigma_i^2 = \cos^2\delta\, \sigma_{\rm t}^2 + \sin^2\delta\,
  \sigma_r^2$  [$\sin\delta = (R_{\odot}/R) \sin\iota$ and $R^2 = 
  R_{\odot}^2 + (zD)^2 - 2 z D R_{\odot} \cos{\iota}$ 
  ($\iota = 82^o$ is the angle between GC and LMC as 
  observed from the Earth)].  Detailed derivations and
  numerical values are given in the Appendix.

As in paper I, 
the halo model corresponding to BHBFS with the power-law 
density profile
$\gamma = 3.4$ and the dispersion given by (\ref{disp.fit}) and Jeans'
equation will be called the `concentrated sphere' (`CS').  
More generally, we will assume for our study that the MHO halo can be 
described by a member of a class of models specified
by five parameters: $\gamma$, $\sigma_{o}$, $\sigma_{+}$ 
$r_o$ and $l$.
For instance, the model
with $\gamma =2$ and constant velocity dispersion $\sigma_{\rm r}
= \sigma_{\rm t} = V_{\rm c}/\sqrt{2}= 156\,$km/s (i.e., 
$\sigma_o = V_{\rm c}/\sqrt{2}$, $\sigma_+ =0$) is 
just the familiar singular
isothermal sphere (SIS).

\section{Distribution of measurable quantities}

\begin{figure*}[t]
\label{distr.CS}
\centering
\hbox{\epsfxsize= 6.5 cm \epsfbox[70 250 350 720]{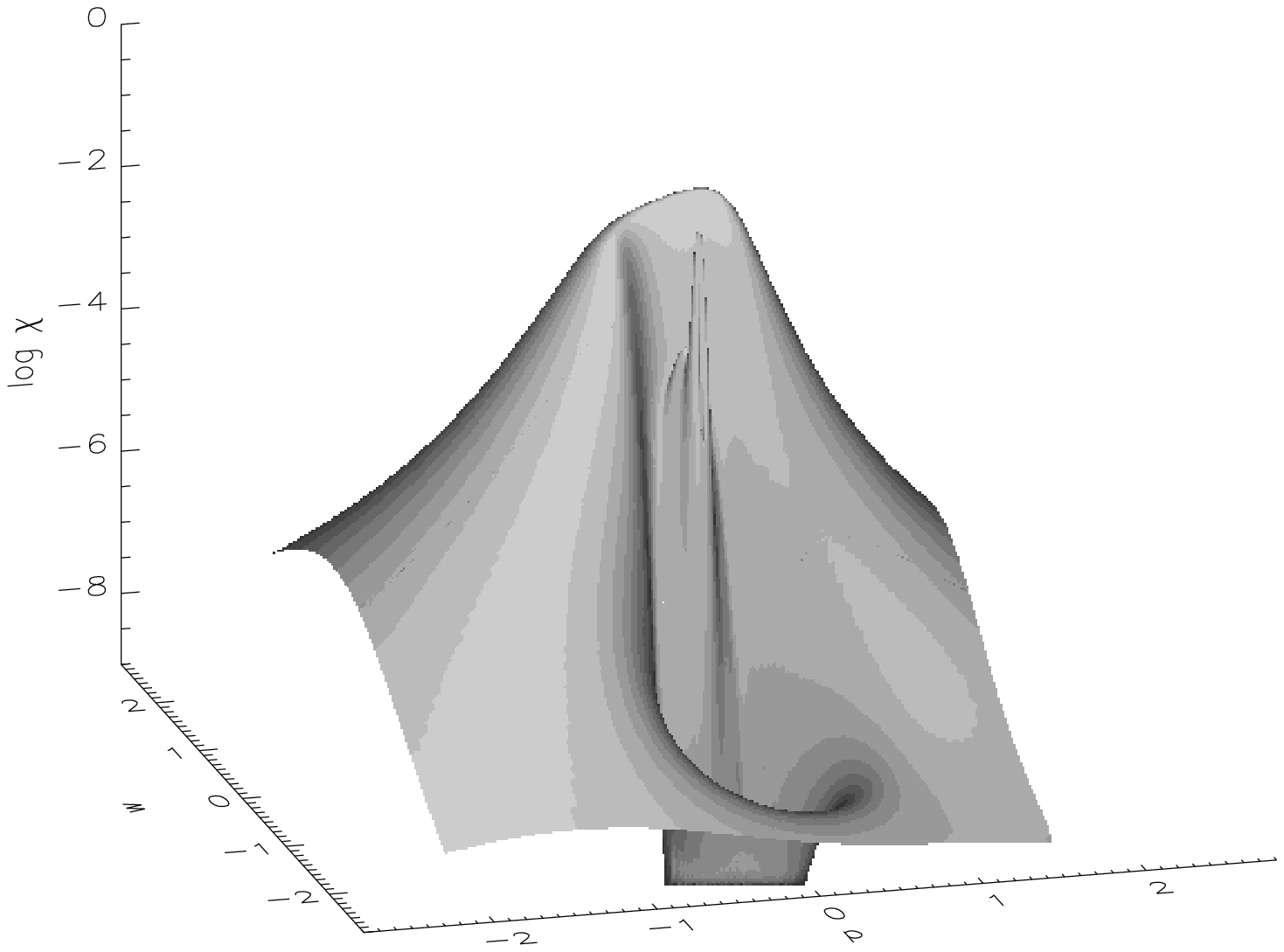}
      \epsfxsize= 5 cm \epsfbox[-280 -170 150 570]{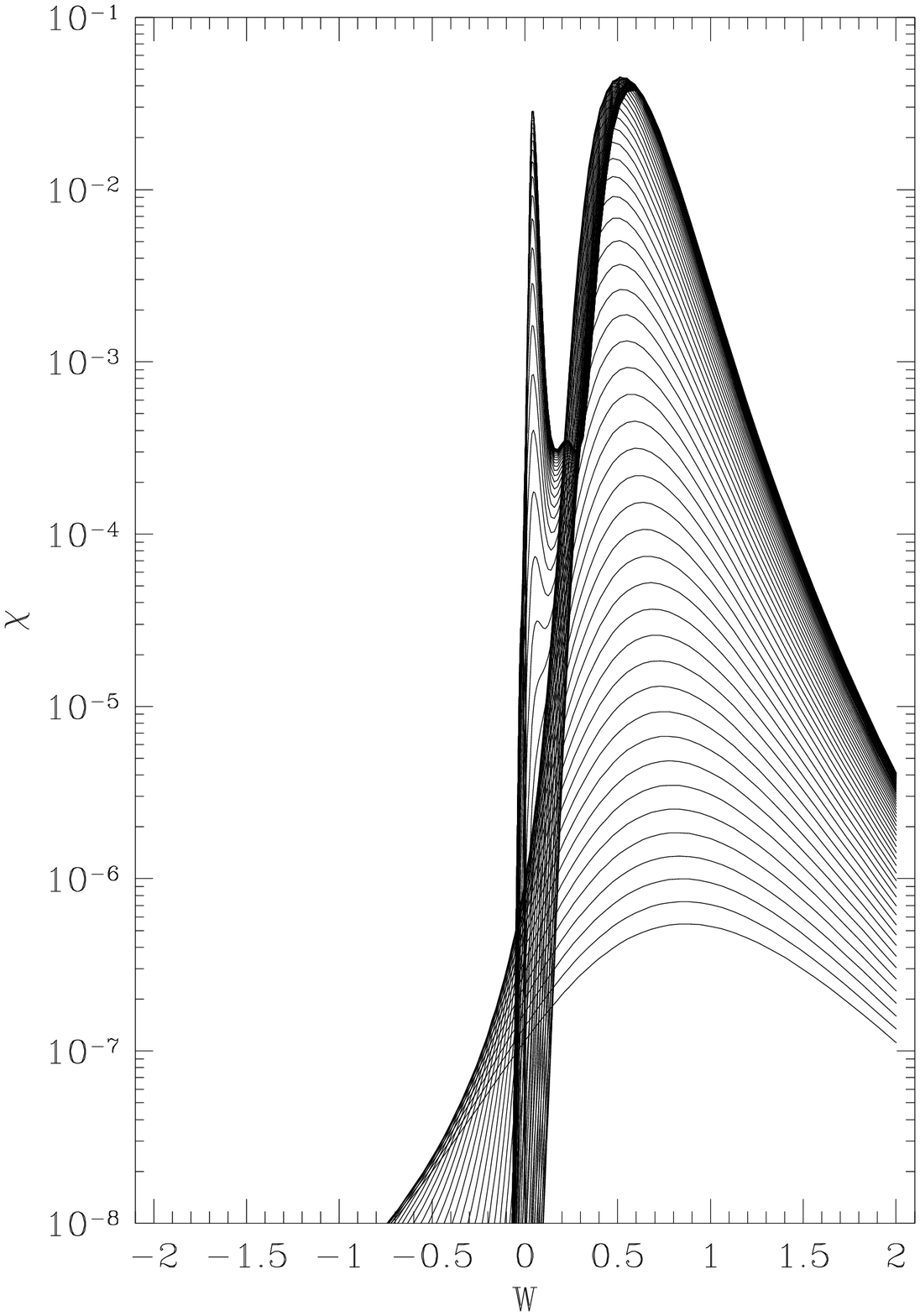}}
\hbox{\epsfxsize= 6.5 cm \epsfbox[70 350 350 600]{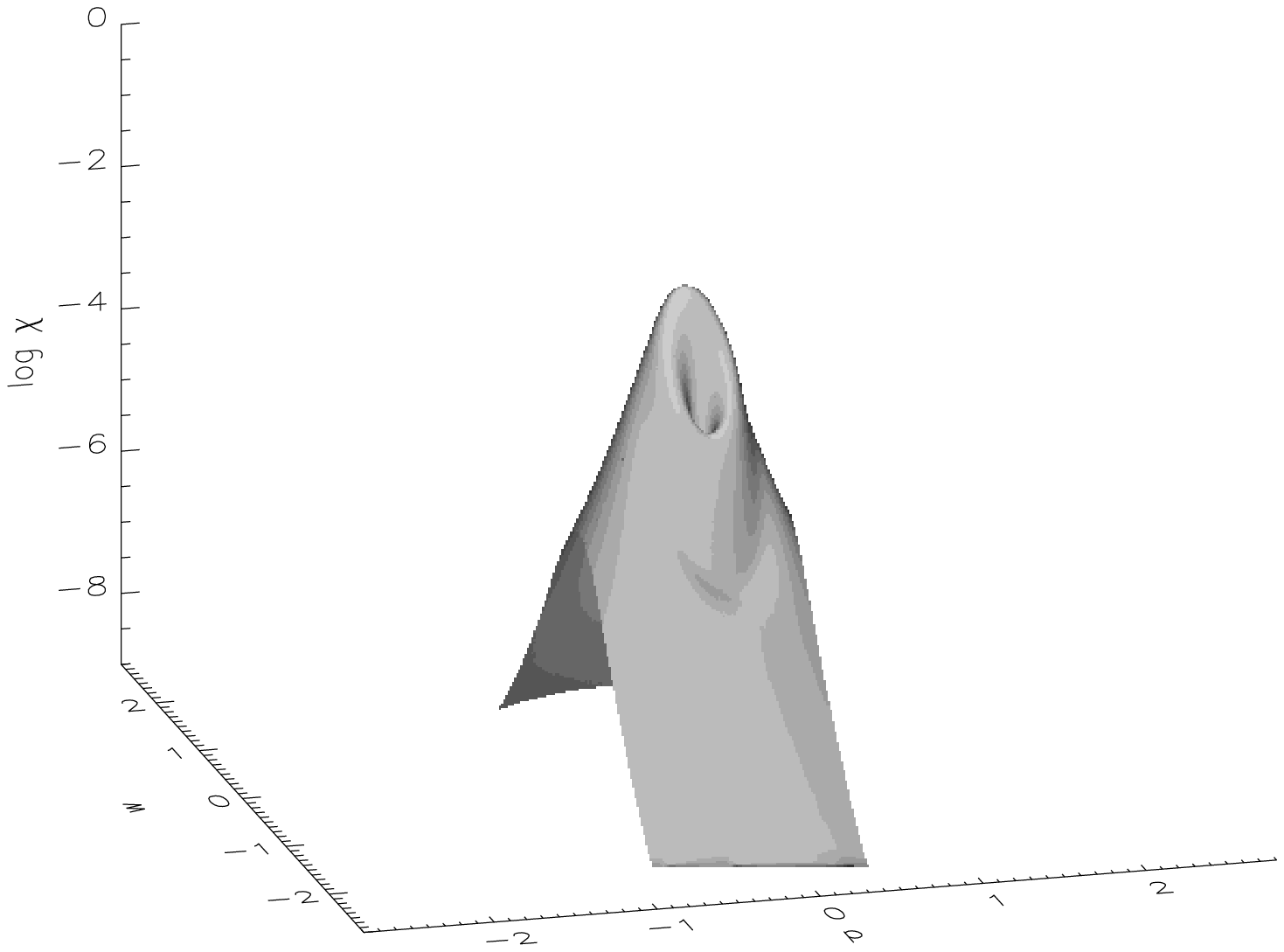}
      \epsfxsize= 5 cm \epsfbox[-280 30 150 220]{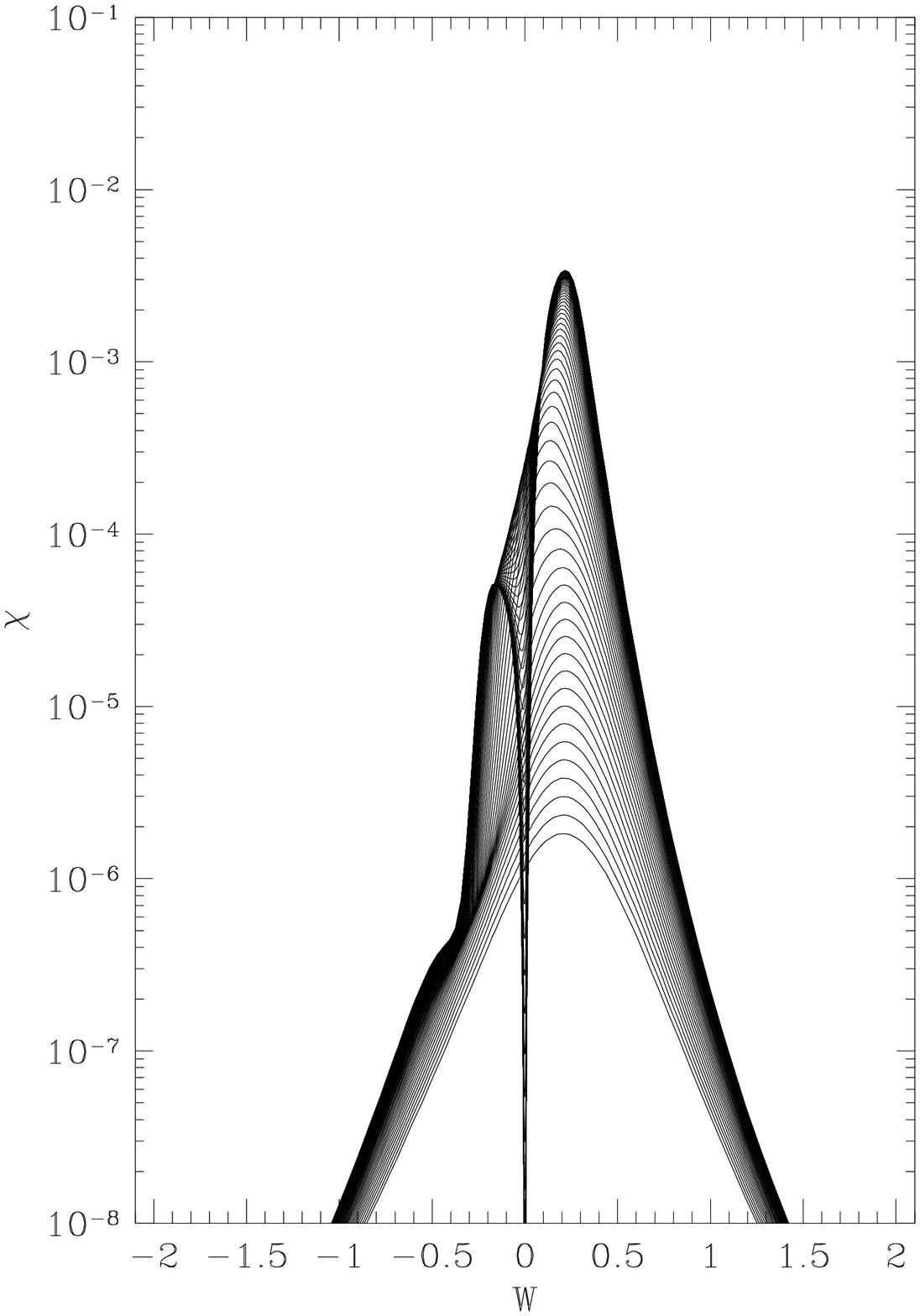}}           
\caption{Plots of $\chi$ with the CS halo (Sun's and LMC motion
taken into account) and a delta mass function
($\mu = 0.4$) at $T=15.1$ (top) and
$T=81.4$ (bottom) days.
}
\end{figure*}

\begin{figure}[t]
\label{dPTu.ps}
\centering
\hbox{\epsfxsize= 9 cm \epsfbox[250 50 650 760]{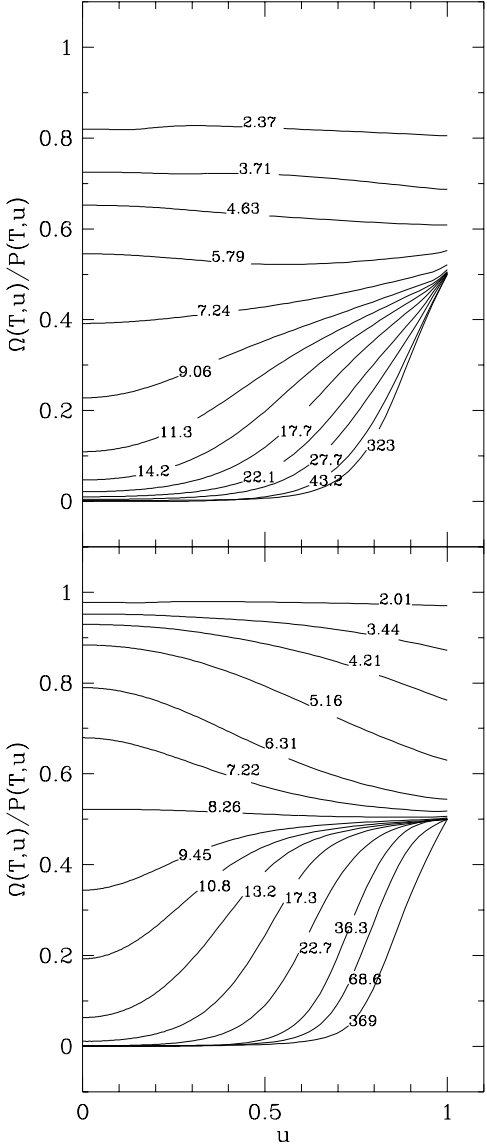}}               
\caption{The fraction $\Omega(T,u)/P(T,u)$ of events detected
in one detector (i.e., with $u < u_{\rm th}=1$) that are {\it not}
detected in the other detector.  The upper panel gives
the plots for the SIS halo model (without Sun's or LMC motion) 
and the bottom for the
CS (with Sun's and LMC motion)
model, both with a delta mass function, $\mu=0.4$. The lines
are labeled with event durations $T$ in days.  The total (i.e.,
integrated over $T$) fraction of {\it single} events
out of all events detected in a detector is 12\% in the upper case
and 20\% in the bottom case:  the CS model gives 
more events at low $z$, where $R_{\rm E}/r(1-z)$ is smaller.
}
\end{figure}

\begin{figure}[t]
\label{dPTu.pic.ps}
\centering
\hbox{\epsfxsize= 10 cm \epsfbox[-20 230 680 600]{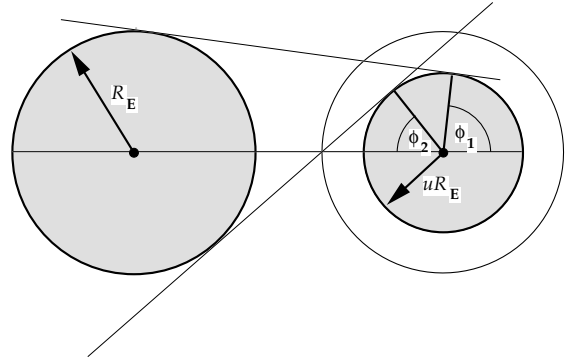}}               
\caption{The lines tangential to both the circles of radius
$R_{\rm E}$ and $u R_{\rm E}$ are drawn to find the sector of
size $2(\phi_1 +\phi_2 )$ from which the events at $u$ ($<u_{\rm th}=1$)
in the right-hand side detector are not detectable by the one
on the left-hand side.  The distance between the centres of the circles
is $r(1-z)$.
}
\end{figure}

\begin{figure}[t]
\label{dTdz.ps}
\centering
\hbox{\epsfxsize= 10 cm \epsfbox[60 30 700 750]{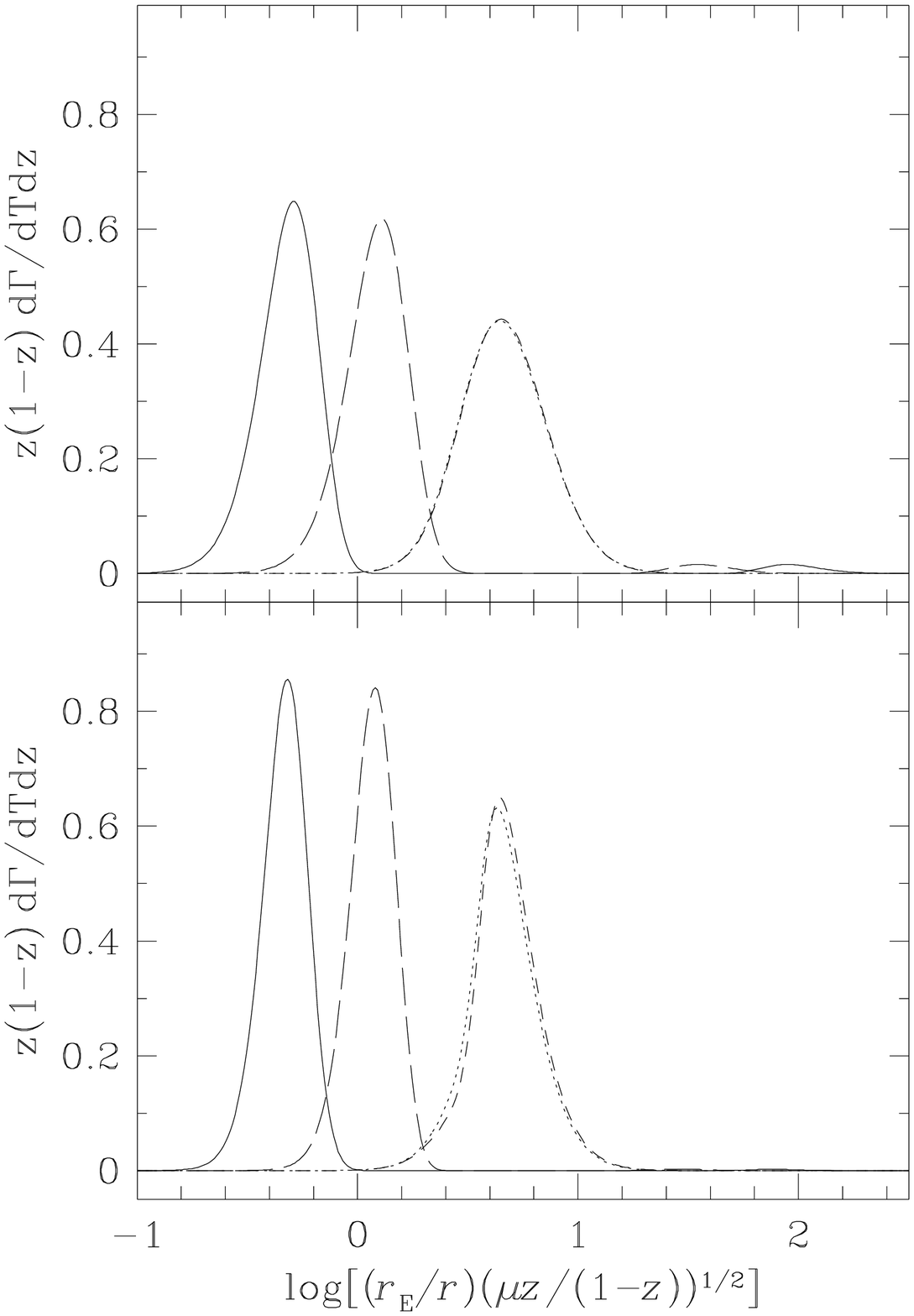}}               
\caption{Normalised (i.e. of a constant area under the curve) 
$d\Gamma/dTdz$ at $T=4.6$ (solid), $T=11.2$ (long dash),
$T=166$ (dotted) and $T=361\,$days (short dash) for the SIS (without 
 Sun's or LMC motion; top)
and CS (with Sun's and LMC motion; bottom) halo models.  
In both cases the mass function
is delta at $\mu=0.4$.  The 166d and 361d curves are virtually 
indistinguishable.
}
\end{figure}

In order to understand  basic features of the distribution functions
$\chi(T,p,w)$ [and consequently $\Psi(T,p,u_1,u_2)$] and $\Omega(T,u)$
we will at first limit ourselves to the relatively
simple case of the SIS halo, neglect the motion of the Sun (i.e.,
the detectors) and the LMC and assume that all MHOs have the
same mass, $\mu =\bar{\mu}$ (see Fig.~4).

With the above assumptions
\begin{eqnarray}
\label{chi.prop}
\chi(T,p,w)\propto \frac{(1-z)^2 z^4 H(z)}{T^4}\exp\left[-\frac{1}{2}
  \frac{r_{\rm E}^2 \mu}{\sigma^2}\frac{z(1-z)}{T^2}\right],
\end{eqnarray}
where $z(p,w)$ is given by equation~(\ref{z.a}).  At  sufficiently small
$a^2 =p^2 +w^2$, a MHO is near LMC, $z\approx 1$, $1-z\approx 
(r_{\rm E}/r)^2 \mu a^2$ and we thus have
\begin{equation}
\label{chi.small.a}
\chi\sim a^4\exp\left[-\frac{1}{2}\left(\frac{r_{\rm E}^2 \mu}{r\sigma T}
 \right)^2 a^2 \right].
\end{equation}
Near the origin of the $p,w$ plane $\chi$ grows at first as 
$a^4$ (from $\chi =0$ at $a=0$) to a maximum value at $a_1 =
2r\sigma T/r_{\rm E}^2 \mu$ (see the `funnel' emerging from $w=0$
on the right of Fig.~5,  which is a projection along
the $p$-axis of the 3D plot on the left), which for the values of
ths SIS model parameters and $\mu=0.4$, $r=2\,$AU takes on value
$a_1 =T/508\,$d.

At large $a$, $z\approx (r/r_{\rm E})^2/\mu a^2$ (close to the detector
at $z=0$) and  thus
\begin{equation}
\label{chi.big.a}
\chi\sim\frac{1}{a^8}\exp\left[-\frac{1}{2}\left(\frac{r}{\sigma T}\right)^2
  \frac{1}{a^2}\right].
\end{equation}
This expression reaches a maximum at $a_2 = (1/2^{3/2}) r/\sigma T = 
7.9\,{\rm d}/T$,
and falls off as $\chi\propto a^{-8}$ for $a >a_2$.

For $T <T_o$, where 
\begin{equation}
\label{To}
T_o \equiv\frac{1}{2^{5/4}}\frac{r_{\rm E}\sqrt{\mu}}{\sigma} =63\,{\rm d},
\end{equation}
we have $a_1 < a_2$ and an annular `valley' will exist between the two
maxima (see the top of Fig.~5) separating the
inner `funnel' from the outer circular wall of the `volcano.'  The detected
events would thus belong to two distinct classes distinguished
by the magnitude of $a$.

As we shift toward larger $T$'s the `valley' turns shallower and flattens
out completely around $T=T_o$.  For $T>T_o$ (see the bottom of 
Fig.~5) only one maximum at $a=a_1$ will persist.

Introduction of the anisotropic velocity dispersion of
the CS halo model will distort the circular contour lines of 
Fig.~5 into elliptical ones.  In addition,  switching on
the motion of the Sun and LMC (Earth's revolution around
the Sun neglected; the Earth-satellite line lies in the ecliptic and
chosen orthogonal to the line of sight;
see the Appendix) will `erode' the outer
wall of the `volcano' asymmetrically (Fig.~6):
more MHOs will cross the 1-2 line of Fig.~1 in the
upward direction.  The motion of the line of sight through the
rest frame of the halo 
will also lead to a general shift of the differential
event rate $d\Gamma/dT$ toward shorter durations (Fig.~4).

In Fig.~7 we plot for various durations $T$ the fraction 
$\Omega(T,u)/P(T,u)$
of events detected in one detector (say, $u\equiv u_2 < u_{\rm th}$) that
are not detectable ($u_1 > u_{\rm th}$) in the other one.
At short $T'$s (few days) the event rate is dominated by MHOs that cross the 
line of sight close to the Sun.  For these MHOs $R_{\rm E} = r_{\rm E}
\sqrt{\mu}\sqrt{z(1-z)}$ is smaller than the projected
separation $r(1-z)$ between the lines 
of sight with respect to the two detectors ($a>1$).
The elementary construction of Fig.~8 shows that
of all MHOs passing at impact parameter $uR_{\rm E}$ relative to
detector 2, the fraction
\begin{eqnarray}
\label{OmP}
\frac{\Omega(T,u)}{P(T,u)} &\approx& \frac{\phi_1 +\phi_2}{\pi} 
      \nonumber \\
      && \hspace{-1.2cm} 
  = \frac{1}{\pi}\left[ \cos^{-1} \frac{R_{\rm E}}{r(1-z)}(1-u)
     + \cos^{-1} \frac{R_{\rm E}}{r(1-z)}(1+u)\right]
     \nonumber \\
     && \hspace{-1.2cm}
     \approx 1 -\frac{2}{\pi} \frac{R_{\rm E}}{r(1-z)} -\frac{1}{3\pi}
      \left[\frac{R_{\rm E}}{r(1-z)}\right]^3 \left(1 + 3\,u^2\right)
      \nonumber \\
\end{eqnarray}
will pass at impact parameters greater than $R_{\rm E}$ 
with respect to detector 1 and thus not be detected if
$u_{\rm th} =1$ (we here assume the 2-dimensional velocity distribution
is isotropic as is the case for the SIS model without the Sun's and
LMC's motion).  

At $R_{\rm E} \ll r(1-z)$ [$z < 1/(1 + \mu r_{\rm E}^2/r^2)\approx 
 (r/r_{\rm E})^2/\mu =0.02$] fraction (\ref{OmP}) depends weakly
 on $u$; the flatness of the corresponding curves can be observed in
 the top part of Fig.~7. Naturally, the $\Omega/P$
 curves are depressed as $T$ [and thus $R_{\rm E}/r(1-z)$] is increased.  

 For MHOs passing at larger $z > 0.02$  (corresponding to typical times
 $T \approx R_{\rm E}/\sqrt{2}\sigma > r/\sqrt{2}\sigma\approx 15\,$d)
 the Einstein radius will  be longer than 
 the projected Earth-satellite 
 distance $r(1-z)$.  This means that an event of
 a small impact parameter
 $u<1 - r(1-z)/R_{\rm E}$, with respect to one detector, will
 inevitably be detected [$\Omega(T,u)/P(T,u)=0$] by the other one.
 For $u > 1 - r(1-z)/R_{\rm E}$, the  fraction detectable in only
 one detector rises approximately
 as 
 \begin{equation}
 \label{OmP.bigT}
 \frac{\Omega(T,u)}{P(T,u)} = \frac{1}{\pi}\cos^{-1} \frac{R_{\rm E}}{r(1-z)}(1-u)
 \end{equation}
 with increasing $u$ and reaches $1/2$ at $u=u_{\rm th}=1$.

The correspondence between the event durations and typical $z$'s is
illustrated in Fig.~9.  The differential rate $d\Gamma/dTdz$
is relatively sharply peaked if plotted with respect to the coordinate
$(r_{\rm E}/r)[\mu z/(1-z)]^{1/2} = R_{\rm E}/r(1-z) =1/a$ instead of $z$
[the area element below a curve in Fig.~9 differs
from $(d\Gamma/dTdz)dz$ only by a constant factor];  we used this fact
in the above discussion to relate a narrow range of $z$ 
(or, rather, simply a
{\it single} value of $z$) to each $T$.

\section{Inference of the MHO mass function and the halo structure}

Even at first sight it seems obvious that 
the increased amount of information
obtained from parallax microlensing measurements should allow a more
reliable determination of the structure of the halo and the MHO
mass function than just the measurements of event durations.
Indeed, the measurable quantity $p\, T = r(1-z)\cos\phi /v_{\rm n}$ 
(equation \ref{p.def}) involves
only the {\it kinematic} properties of a MHO 
(i.e., excluding its mass).  In the case of
resolved parallaxes, an additional quantity, $w$, is measured
and this gives us the 2-dimensional velocity projected on the
observer's plane (the so-called {\it reduced} velocity; see. e.g., 
Gould 1994b) 
\begin{equation}
\label{reduc.vel}
\tilde{\bf v}\equiv \frac{1}{1-z}{\bf v}_{\rm n} =
\frac{r}{T}\frac{\bf a}{a^2} = \frac{r}{T}\frac{1}{p^2 + w^2}(p, w).
\end{equation}
Although the 4-fold ambiguity of the degenerate parallaxes reduces the
quality of available information, both types
of parallax microlensing would constrain the halo independently 
of the MHO masses; this 
should make it possible to separate the effects of the masses
and determine the MHO mass function.

\begin{figure}[t]
\label{errors.mf}
\centering
\hbox{\epsfxsize= 10 cm \epsfbox[270 40 700 750]{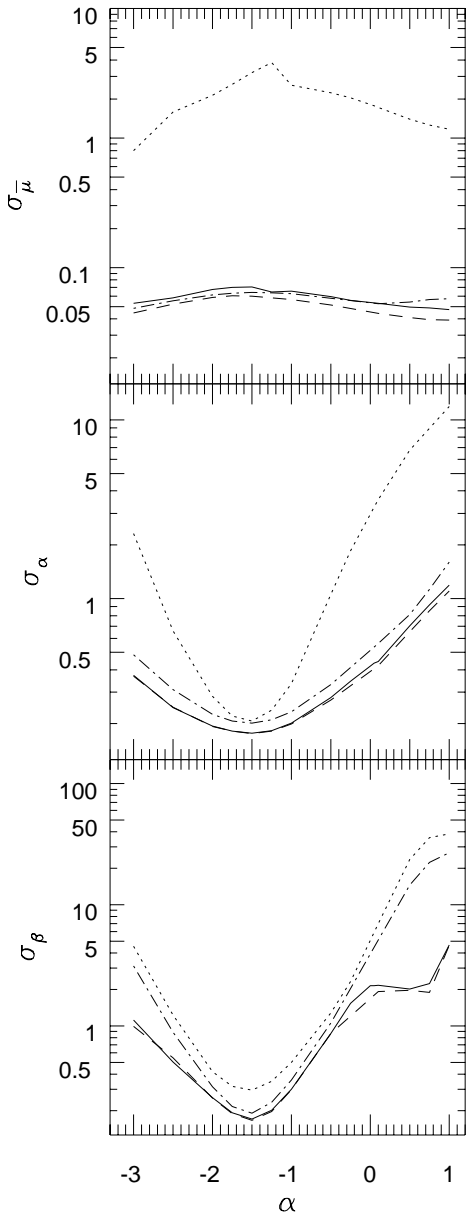}}
\caption{The Cramer-limit errors with an unknown (variable) halo
from resolved parallaxes (solid lines) and event durations only (dotted).
For a fixed CS halo model, the errors are shown as dashed lines
(resolved parallaxes) and dot-dashed (event durations only) lines.
All errors are given for $\bar{\mu}=0.4$, $\beta =2$ and $N = 100$ events.}
\end{figure}

In order to assess the information gain due to parallax microlensing
we will investigate simulated maximum likelihood inference of the five 
parameters ($c_1 \equiv \gamma$, $c_2 \equiv \sigma_o$,
$c_3 \equiv \sigma_+$, $c_4 \equiv r_o$, $c_5 \equiv l$) of the halo model
and the three parameters ($c_6 \equiv \bar{\mu}$, $c_7 \equiv \alpha$,
$c_8 \equiv \beta$) of the mass function based on the normalised probability
distributions $\hat{\chi}(T,p,w)$, $\hat{\Omega}(T,u)$ 
(equation \ref{resolv.norm}) for resolved or
$\hat{\Psi}(T,p,u_1, u_2)$, $\hat{\Omega}(T,u)$ (equation \ref{degen.norm})
for degenerate parallax measurables.

Ideally, one would prefer to obtain an answer through a series of 
Monte-Carlo simulations, but the enormous numerical task of computing
the above 3-dimensional distribution $\chi$ for many points in the
8-dimensional parameter space compels us to seek more economical 
alternatives.  For a sufficiently large number $N$ of detected events,
the average errors $\sigma_{\mu}\equiv 
\left\langle [c_{\mu} - c_{\mu}^{(o)}]^2 \right\rangle ^{1/2}$
of the parameters ($c_{\mu}^{(o)}$ is the `real' value of a parameter)
can be shown (see, e.g., paper I) to approach asymptotically
the Cramer limit
\begin{equation}
\label{cramer.limit}
\left\langle\left(c_{\mu} - c_{\mu}^{(o)}\right) 
\left(c_{\nu} - c_{\nu}^{(o)}\right)\right\rangle \longrightarrow
{\cal C}_{\mu\nu},
\end{equation}
where ${\cal C}_{\mu\nu}$ is the inverse of the information matrix
\begin{eqnarray}
\label{info.matr}
I^{(N)}_{\mu\nu} &=& N\int\int\int\int dTdPdu_1du_2 \frac{\hat{\Psi}_{,\mu}
      \hat{\Psi}_{,\nu}}{\hat{\Psi}}
       \nonumber \\ 
     & &  + \; 2 N \int\int dTdu \frac{\hat{\Omega}_{,\mu}
      \hat{\Omega}_{,\nu}}{\hat{\Omega}}
\end{eqnarray}
($_{,\mu}$ denotes derivative with respect to the $\mu$'th parameter)
for degenerate parallaxes and by analogy for the resolved parallaxes.

In our numerical experiment we assume that the parameters
of the concentrated sphere
(CS)  describe accurately the MHO halo.  The Cramer errors of the 
inference of the mass-function
parameters for $N=100$ are shown in Fig.~10 as functions of the
mass-function slope $\alpha$. The resolved 
parallax errors obtained with the assumption 
that both  the halo structure and the mass function parameters
were unknown (and thus variable in the maximum likelihood fitting)
are given as solid lines.  For comparison, we also show the
Cramer-limit errors (dotted lines) for measurements of event
durations only.  In addition, we plot the resolved parallax errors 
(dashed lines) and the errors based on event durations only
(dot-dash) assuming that the halo parameters are known precisely
(and thus not varied for the maximum likelihood fit).  In all computations
we take into account the motion of the Sun and LMC,
assume $r=2$ and that the Earth-satellite segment is in the
ecliptic and orthogonal to the Earth-LMC line of sight 
(see the Appendix). 

The most striking feature of these plots is a significant
reduction of the error in $\bar{\mu}$ brought about by parallax
measurements.
As shown in paper I, by changing the parameters of the halo model one
can match closely event duration distribution curves corresponding
to widely different average masses.  This is reflected in the
large errors in $\bar{\mu}$ if the halo parameters are allowed to vary
in addition to the parameters of the mass function.
The extra information provided by parallaxes effectively allows us
to constrain the halo as to bring $\sigma_{\bar{\mu}}$ down to the values 
comparable to those of inference with a
fixed (i.e., `known') halo model. (Notice that the 
gain in accuracy due to parallaxes for a known halo structure is
modest by comparison.) 

\begin{figure*}[t]
\label{corr.plot}
\centering
\hbox{\epsfxsize= 8 cm \epsfbox[40 320 500 750]{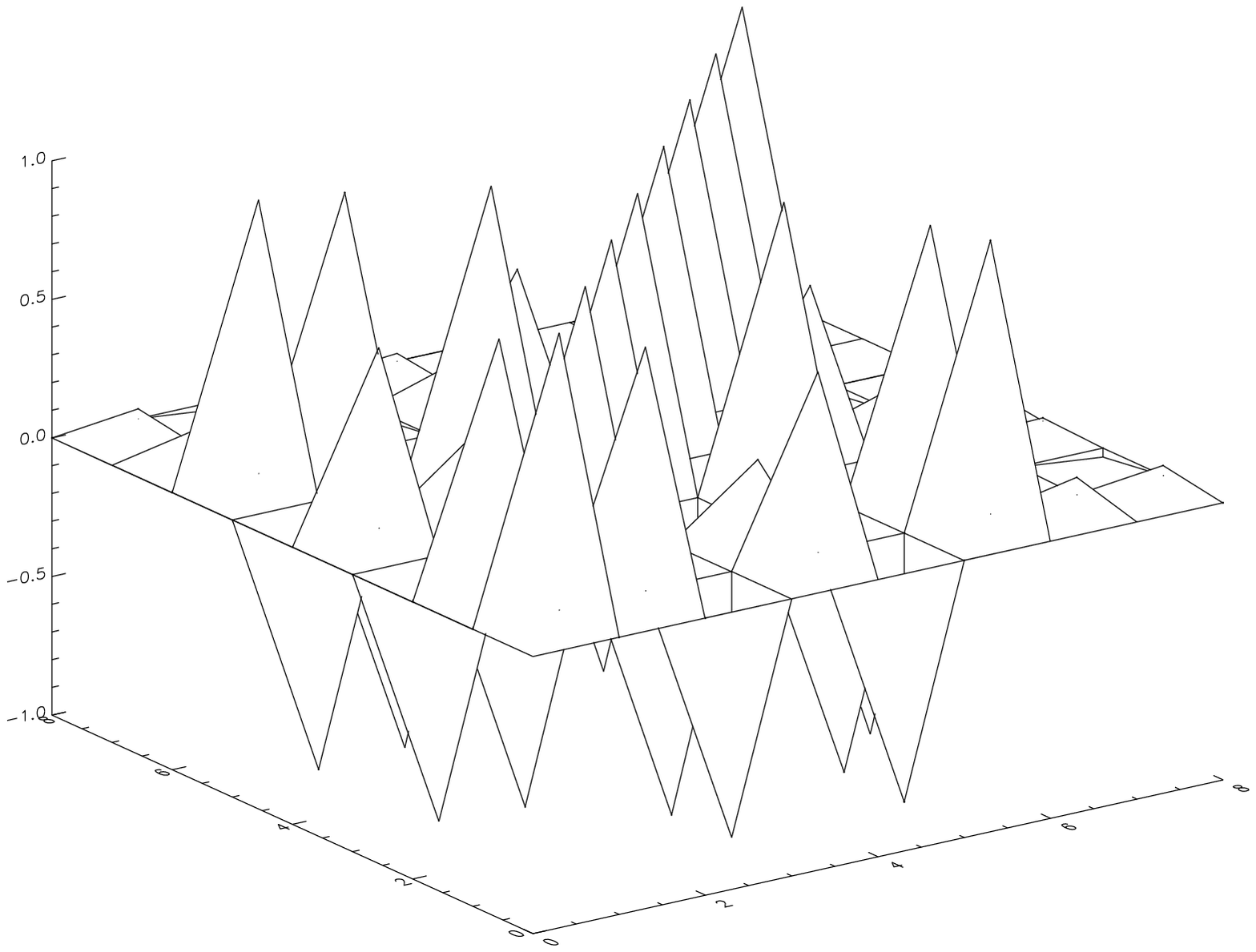}
      \epsfxsize= 8 cm \epsfbox[500 320 960 750]{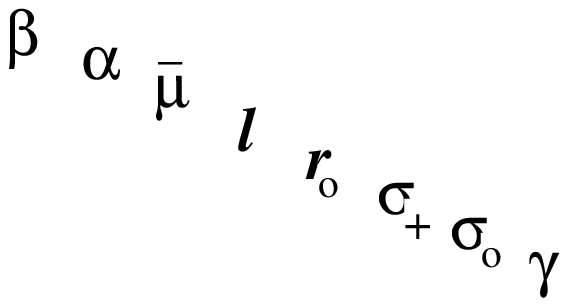}
      \epsfxsize= 8 cm \epsfbox[500 320 960 750]{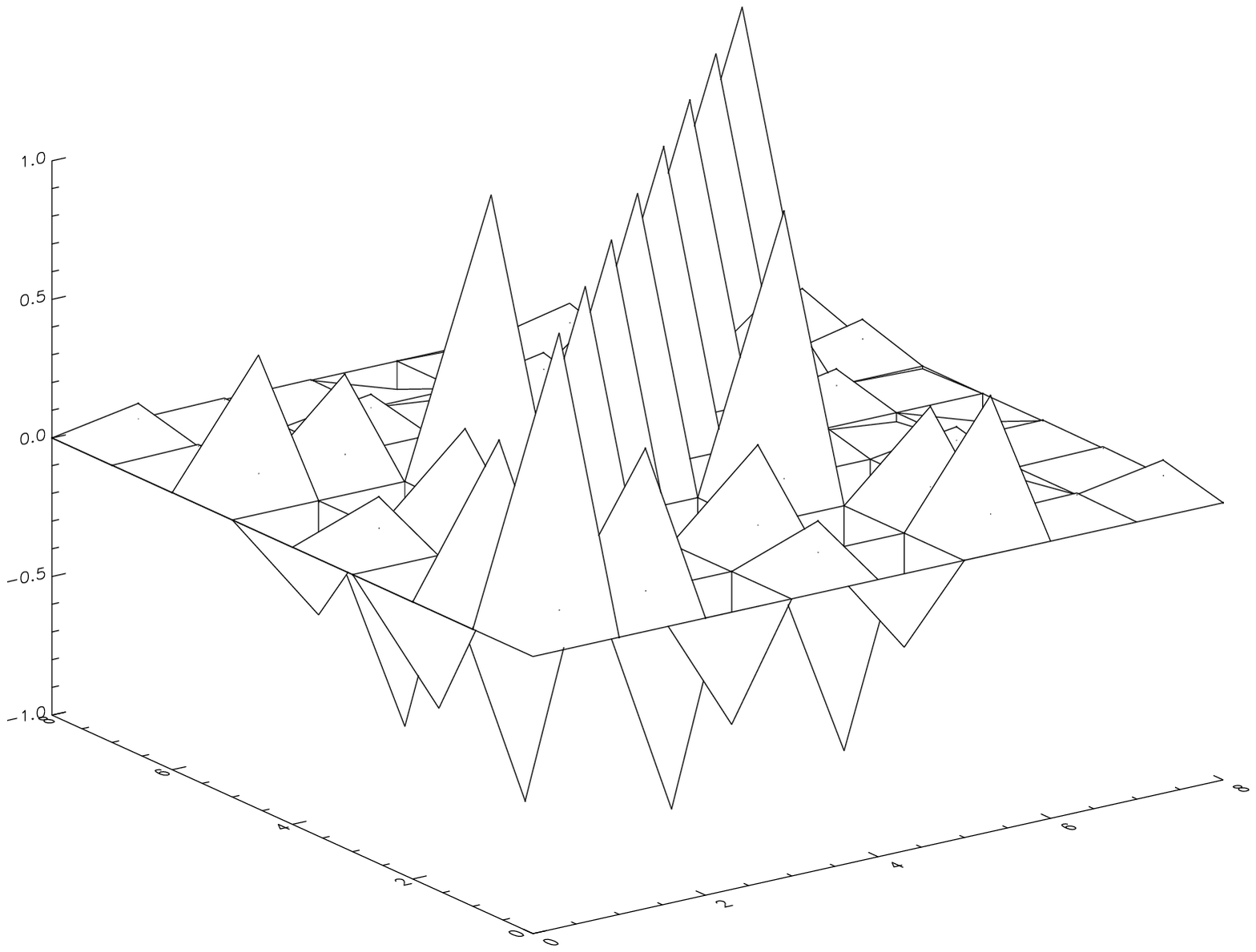}}
\caption{Correlation matrices for Cramer errors of inference based on
event durations only (left) and resolved parallaxes (right).  The
underlying halo model is CS and the parameters of  the mass  function
function are $\bar{\mu} =0.4$, $\alpha =-1.5$ and $\beta = 2$. 
}
\end{figure*}

In paper I we have concluded that for a certain range of $\alpha$
(very roughly, $-2\alt\alpha\alt 0$) the inference of
$\alpha$ and $\beta$ is relatively weakly affected by the uncertainty
of the halo structure.  This again is manifested in errors of 
the same (small)
order near $\alpha=-1.5$  in all four cases shown in Fig.~10.
However, away from $\alpha=-1.5$, $\sigma_{\alpha}$ and $\sigma_{\beta}$
grow large for event duration-based inference.
Here again, the parallaxes restrain this growth significantly (especially for 
$\sigma_{\alpha}$) and keep the errors down  in the range characteristic of
the fixed-halo inference.
Interestingly, degenerate parallax errors $\sigma_{\bar{\mu}}$
 $\sigma_{\alpha}$ and $\sigma_{\beta}$ are only a few percent smaller 
 than the resolved parallax values (and we omit the
 corresponding plots).   
 
 As suggested above, the significant
 improvement in the accuracy of the mass function determination should
 be ascribed to the  disentanglement of the halo structure
 from the MHO masses.  This disentanglement is obvious
 from the plots in Fig.~11 of the correlation matrix
 \begin{equation}
 \label{corr.matr}
 \frac{{\cal C}_{\mu\nu}}{ \sqrt{{\cal C}_{\mu\mu}} 
                   \sqrt{{\cal C}_{\nu\nu}}}.
 \end{equation}
 The off-diagonal spikes `coupling' the halo with
 the mass function are noticeably less pronounced in the
 parallax case (right).
 [Notice, e.g., the strong correlation between $\bar{\mu}$ and
 $\gamma$ in the duration based inference (left) and its 
 significant drop in the  parallax case (right).]

\begin{figure}[t]
\label{errors.halo}
\centering
\hbox{\epsfxsize= 10 cm \epsfbox[150 60 530 750]{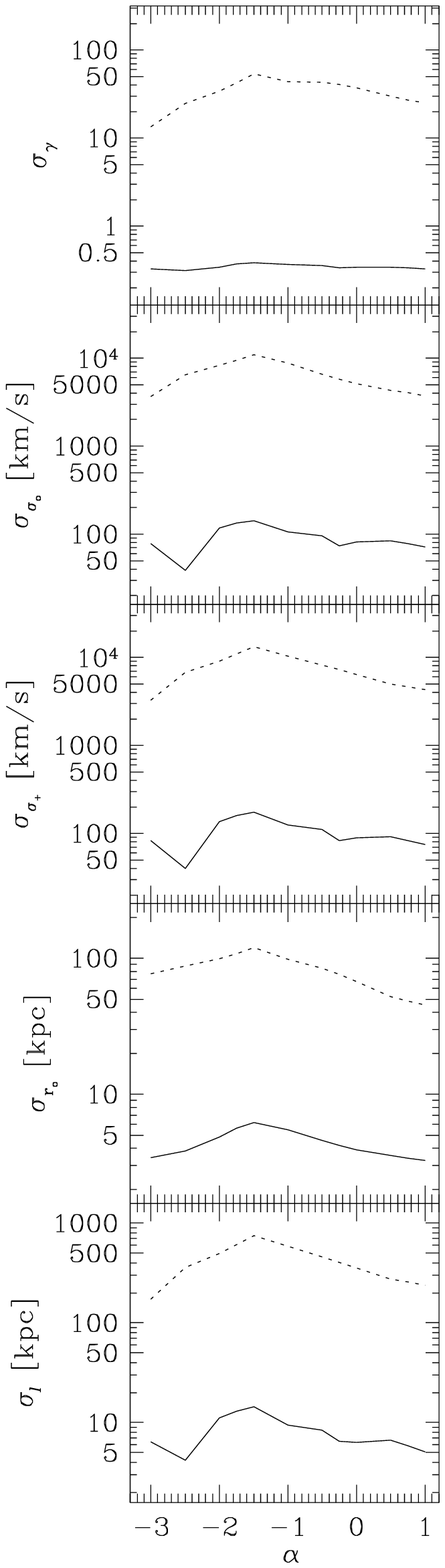}}
\caption{The Cramer-limit errors (at $\bar{\mu}=0.4$, $\beta=2$ and
$N = 100$) 
 of the halo-model parameters from resolved parallaxes 
 (solid lines) and event durations only
 (dotted lines).}
\end{figure}

Further comparison of the two plots of Fig.~11 also shows
 a significant reduction in the correlation among some of the halo
 parameters due to the parallax measurements.  Does this imply that
 the halo structure itself could be inferable?  Figure~12
 indeed displays a remarkable suppression of the Cramer errors
 of the halo parameters' inference: the (resolved) parallax errors 
 at least are 
 of the same order of magnitude as the parameters  themselves
 or smaller.\footnote{
 As discussed in paper I, the Cramer errors can serve as reliable estimates
 of actual errors only if they are small in comparison with the 
 corresponding parameters that one is trying to infer.  Typically,
 we would expect the nonlinear dependence of the distribution function
 of the observables on the underlying parameters to make the actual
 errors smaller that the Cramer estimates when these are relatively large.
 In this case the actual errors would fall slower than $N^{-1/2}$ and
 approach the Cramer limit from {\it below}.}  In particular,
 the density profile index $\gamma$ can be determined rather
 accurately.  [The difference between
 the degenerate and resolved parallax errors is more pronounced for
 the halo model parameters than for the mass function; still, the
 degenerate parallax errors are larger than the resolved parallax
 ones by at most 20\%.  The measurement of $p$ and the 
 information contained in $u_1$, $u_2$ and $\Omega(T,u)$ are
 thus sufficient to constrain the halo even if the determination of
 $w$ is subject to the 4-fold  ambiguity.]

\begin{table*}
  \begin{center}
    \begin{tabular}{cccccccccc}\hline
    $\sigma_{(1)} = 0.033$,  & 
    	  ${\bf V}_{(1)}$ = & ( 0.7667, & 0.2005, & 0.5791, & 0.1562,
	            & -0.1037, & 0.0233, & 0.0149, & -0.0258 ) \\
    $\sigma_{(2)} = 0.795$,  &
          ${\bf V}_{(2)}$ = & ( 0.0362, & 0.7289, & -0.0838, & -0.5037, 
	            & 0.4541, & 0.0229, & -0.0048, & 0.0035 ) \\
    $\sigma_{(3)} = 0.126$,  &
          ${\bf V}_{(3)}$ = & ( -0.4744, & 0.2243, & 0.4946, & 0.1677, 
	            & -0.0685, & 0.3439, & -0.5733, & 0.0175 ) \\
    $\sigma_{(4)} = 0.383$,  &
          ${\bf V}_{(4)}$ = & ( 0.0511, & 0.2676, & -0.3075, & 0.8129,
	            & 0.4111, & 0.0083, & -0.0103, & -0.0338 ) \\
    $\sigma_{(5)} = 3.316$,  &
          ${\bf V}_{(5)}$ = & ( -0.0158, & -0.5050, & 0.3601, & -0.0903,
	            & 0.7789, & -0.0143, & -0.0019, & -0.0001 ) \\
    $\sigma_{(6)} = 0.182$,  &
          ${\bf V}_{(6)}$ = & ( 0.3327, & -0.1966, & -0.3563, & -0.1221,
	            & 0.0442, & 0.8117, & -0.2121, & 0.0540 ) \\
    $\sigma_{(7)} = 0.102$,  &
          ${\bf V}_{(7)}$ = & ( -0.2675, & 0.1085, & 0.2431, & 0.0859,
	            & -0.0271, & 0.4702, & 0.7796, & -0.1434 ) \\
    $\sigma_{(8)} = 0.081$,  &
          ${\bf V}_{(8)}$ = & ( -0.0270, & 0.0343, & 0.0510, & 0.0499,
	            & 0.0048, & 0.0186, & 0.1351, & 0.9871 ) \\
         \hline
    \end{tabular}
  \end{center}
\caption{Eigen-errors (for $N=100$ events)
and the corresponding eigen-vectors of
the relative error matrix for the CS halo and $\bar{\mu}=0.4$,
$\alpha =-1.5$ and $\beta =2$.} \label{eigenvec}
\end{table*}

\begin{figure}[t]
\label{halo.modes}
\centering
\hbox{\epsfxsize= 10 cm \epsfbox[30 30 660 750]{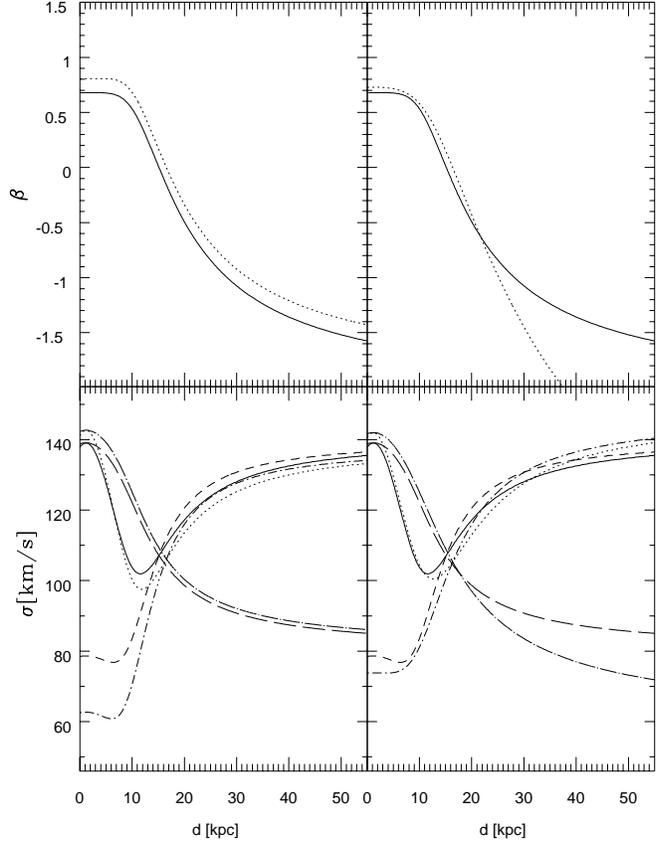}}
\caption{Comparison of the lowest-error halo mode, ${\bf V}_{(1)}$, at
  $\epsilon = 0.05$ (left) and
  the highest-error, ${\bf V}_{(5)}$ at $\epsilon = 0.5$ (right) halo mode.
  In the bottom plots $\sigma_{\rm r}$ 
  of the CS model is given as long dashed lines
  while its `shifts' corresponding to $\epsilon {\bf V}_{(1)}$
  or $\epsilon {\bf V}_{(5)}$ are given as dot-long dashed lines.  The $\sigma_j
  = \sigma_{\rm t}$ lines are short dashed (CS) and dot-short dashed
  (shifted from CS).  The $\sigma_i$ lines are solid (CS) and dotted
  (shifted).  In all cases $\bar{\mu} = 0.4$, $\alpha =-1.5$ and $\beta =2$.
  }
\end{figure}

The residual but stil considerable correlations among the errors of
the halo parameters indicate that some combinations (functions)
of them (together with the mass function parameters)
 may be possible to determine with a significantly higher accuracy.
In order to investigate this possibility we examine the eigen-values and
eigen-vectors of the {\it relative} error matrix
\begin{equation}
\label{rel.error}
\Delta_{\mu\nu} \equiv \frac{{\cal C}_{\mu\nu}}{c_{\mu}^{(o)}c_{\nu}^{(o)}}
 = \left\langle \left(\frac{c_{\mu}}{c_{\mu}^{(o)}} - 1\right)
           \left(\frac{c_{\nu}}{c_{\nu}^{(o)}} - 1\right) \right\rangle,
\end{equation}
i.e., those normalised vectors ${\bf V}_{(\rho)}\;$  ($ V_{(\rho)}^{\mu}
V_{(\rho)}^{\mu} =1$, using Einstein's convention for summation
over the repeated index $\mu$) that satisfy the eigen-equation
\begin{equation}
\label{eigen.equ}
\Delta_{\mu\nu} V_{(\rho)}^{\nu} = \sigma_{(\rho)}^2 V_{(\rho)}^{\mu}.
\end{equation}
These eigenvectors correspond to the linear expansions of the
desired functions of the parameters in the vicinity of $c_{\mu}^{(o)}$;
they indicate the mutually independent (uncorrelated) infinitesimal
displacements in the parameter space caused by statistical errors of
parallax microlensing detections.  In a certain sense, they are 
the {\it eigen-modes} of
the halo model (plus the mass function) as viewed through the `instrument'
of parallax microlensing.

We illustrate the above point by a specific example.  
Table~1 contains the square roots (`eigen-errors') of
the eigen-values  along with the eigen-vectors of the relative error matrix, 
assuming again the CS halo model and $\bar{\mu}=0.4$, 
$\alpha =-1.5$ and $\beta =2$ (same as for Fig.~11).  
We immediately notice that the eigen-errors for the eigen-`modes' are small
($\sigma_{(\rho)} < 1$) except for a single
mode ($\sigma_{(5)} = 3.316$).  This, the 5'th mode `mixes' rather strongly
$c_2 =\sigma_o$, $c_5 = l$ and to a lesser degree
$c_3 =\sigma_+ $ (see the corresponding components of 
${\bf V}_{(5)}$).  Its mass function components,
$V_{(5)}^6$, $V_{(5)}^7$ and $V_{(5)}^8$, are very small which explains why
its large associated eigen-error does not lead to
large errors of the mass-function parameters.  On the other hand, although
there is an eigen-vector, ${\bf V}_{(3)}$, that strongly mixes the halo
and the mass mass function parameters, its relatively small eigen error, 
$\sigma_{(3)} = 0.126$ contributes little to the errors
of the mass function parameters and thus gives at most moderate correlations
between these and the halo-model errors.

In Fig.~13 we compare changes in the halo model caused
by a displacement in the direction of the vector $V_{(1)}$ of a small
eigen-error with a displacement along $V_{(5)}$. Notice that a much larger
displacement ($\epsilon =0.5$) is needed along $V_{(5)}$ to cause a
discernible effect.  In addition, these changes tend to be concentrated at larger
distances, where the event rate (recall $\gamma =3.4$) is smaller; microlensing
is, not surprisingly, relatively insensitive to such displacements along $V_{(5)}$
in the parameter space.  As a consequence, the large associated
eigen-error, $\sigma_{(5)}$, is the leading culprit for the relatively
large errors of  $c_2 =\sigma_o$, $c_5 = l$ and
$c_3 =\sigma_+ $ evident in Fig.~12.
By contrast, parallax microlensing constrains
fairly well the displacements along the other directions.  In particular,
the eigen-modes {\it closest} to the parameters of the mass function
($V_{(6)}$, $V_{(7)}$ and $V_{(8)}$) all have rather small eigen-errors.

The coefficient $\epsilon =0.5$ used in Fig.~13 for the displacement 
along $V_{(5)}$ is considerably smaller than the associated 
eigen-error $\sigma_{(5)}=3.3$ (see Table~1); the relatively
small $\epsilon$ allows us to stay near the `linear' range
of shifts in the structure of the halo (the range 
of the validity of the Cramer errors).  At $\epsilon =3.3$ the structure
would be changed radically, far beyond what one might expect
on the basis of small linear displacements.  This 
likely implies that actual errors would be smaller
than the Cramer-limit estimate.  The results of such
 strongly {\it non-linear} shifts can be studied only by means of
 Monte-carlo simulations which are beyond the scope
 of this paper.  

The CS halo discussed so far can be regarded as a typical
member of our class of spherical models.  On the other hand,
SIS is peculiar in that the parameters $r_o$ and $l$ can be 
arbitrary as long as $\sigma_+ =0$.  Might this `degeneracy' lead
to a large uncertainty in the inference of the halo structure?

To explore this issue we assume $\gamma=2$, $\sigma_o =156\,$km
$\sigma_+ =10\,$km, $r_o=10.5\,$kpc and $l=5.5\,$kpc, only slightly
deviating from SIS.  The root-quadratic dependence of $\sigma_{\rm r}$
on $\sigma_o$ and $\sigma_+$  (equation \ref{disp.fit}) suggests
we use the squares $c_2 = \sigma_o^2 = 2.4\times 10^4\,$km$^2$/s$^2$
and $c_3 = \sigma_+^2 = 100\,$km$^2$/s$^2$ among the
 parameters whose inference we estimate in the Cramer limit.
 
 At $N=100$ events (we take $\bar{\mu}=0.4$, $\alpha=-2$ and $\beta=2$)
 the resolved parallax errors for the density profile, $\sigma_1 =0.36$,
 and the mass-function parameters, $\sigma_6 = 0.08$, 
 $\sigma_7 = 0.20$, $\sigma_8 = 0.27$, are again small,
 while the errors of the velocity-dispersion parameters
 are much larger: $\sigma_2 = 2.9\times 10^4\,$km$^2$/s$^2$,
 $\sigma_3 = 8.1\times 10^4\,$km$^2$/s$^2$,
 $\sigma_4 = 3.4\times 10^3\,$kpc and $\sigma_5 = 4.7\times 10^3\,$kpc
 (the degenerate parallax errors are, again, only insignificantly
 larger).
 
 The diagonalisation procedure used above for the CS case gives for the
 eigenvectors mixing the velocity-dispersion parameters 
 $\sigma_{(2)}=0.30$, $\sigma_{(3)}=207$, $\sigma_{(4)}=183$
 and $\sigma_{(5)}=1100$.  Although the three eigen-errors
 $\sigma_{(3)}$, $\sigma_{(4)}$ and $\sigma_{(5)}$ seem inordinately
 large, the corresponding shifts $\sigma_{(i)}\bf{V}_{(i)}$ in the 
 parameter space do not produce dramatic changes in the structure 
 of the halo.  Table~2 gives the values of the halo parameters
 after these shifts, as well as the radial and tangential velocity
 dispersions.   Not surprisingly, in all three cases the very large 
 magnitudes of $r_o$ and $l$ ensure position-independent velocity
 dispersion profiles and through very different combinations of
 $\sigma_o$ and $\sigma_+$ lead to $\sigma_{\rm r}$ and $\sigma_{\rm t}$
 relatively close to those of SIS.  
 This `degenerate' (almost) SIS structure shows 
 even more clearly than the above CS-based example
 that large
 uncertainties in the values of the halo parameters are not incompatible
 with a good grasp of the structure of the halo:  parallax
 microlensing can probe  the halo while
 ignoring the vagaries of our parametrisation of its structure.

\begin{table}
  \begin{center}
    \begin{tabular}{cccc}\hline
        & $\sigma_{(3)}{\bf V}_{(3)}$  
         & $\sigma_{(4)}{\bf V}_{(4)}$ & $\sigma_{(5)}{\bf V}_{(5)}$ \\ 
	 \hline
       $\gamma$ & 2.02  &  1.81  & 1.77 \\
       $\sigma_o\,[{\rm km/s}]$& 170  &  116  & 15 \\ 
       $\sigma_+\,[{\rm km/s}]$& 81  &  111  & 272 \\
       $r_o\,[{\rm kpc}]$& $-1.7\times 10^3$ & $-1.1\times 10^3$ &
           $-2.5\times 10^3$ \\
       $l\,[{\rm kpc}]$& $-0.6\times 10^3$ & $-0.5\times 10^3$ &
           $4.3\times 10^3$ \\
	 \hline
       $\sigma_{\rm r}\,[{\rm km/s}]$& 186  &  122  & 158 \\ 
       $\sigma_{\rm t}\,[{\rm km/s}]$& 155  &  160  & 164 \\
 \hline
    \end{tabular}
  \end{center}
\caption{Halo structure parameters and radial and tangential
velocity dispersions after displacements $\sigma_{(i)}{\bf V}_{(i)}$
in the parameter space.
}
\end{table}    

\section{Conclusion}

The use of parallax microlensing for the class of models discussed in 
this paper indeed brings a great advantage:  it allows an effective 
disentanglement of the mass function of the MHOs and the
structure of the halo.  There indeed seems to be a way to go beyond
the somewhat pessimistic conclusions of paper I and infer accurately the
average mass of the MHOs (together with the shape of the mass function)
after detecting a realistic number, $N\sim 100$, of events.  In addition,
one can constrain the halo structure much more tightly than it is possible
through measurement of event durations only.  Moreover, some combinations
of parameters describing the halo structure can be inferred rather accurately.
Virtually all the uncertainty regarding the halo model is then localised
in a few (precisely {\it one} in the CS-based example discussed in Section 5)
`eigen-modes' of the halo, i.e., those displacements in the halo parameter
space that are left `loose' by parallax microlensing while allowing other
displacements to be independently (and tightly) constrained.
These poorly constrained modes correspond to particularly small
changes in the {\it actual} structure of the halo as given by density
and velocity dispersion profiles.  Unfortunately, their
large associated eigen-errors may push us into a non-linear
regime (inadequately charted by the Cramer limit) of 
deviations in the halo structure that can be properly explored only
by time-consuming Monte-Carlo simulations.

Although obtained on the basis of a limited class of
halo models, the above conclusions should be relevant in 
a broader context. The halo structure/mass function disentanglement
(and consequently accurate mass determination) as well as
the inference of some properties of
the halo structure itself should result from the significant 
enhancement of information (as elaborated in Section 5)
due to parallax microlensing even if one allows for a much wider range 
of halo structures.  
For instance, in the more general case of non-spherical haloes,
one may hope to constrain the density and velocity dispersions {\it
along the line of sight} well enough to determine the mass function.
The `uncertain' modes would then describe possibly large
changes in  the halo structure away from the line of sight.
Only parallax microlensing observations in several directions
would presumably suffice to infer the overall structure of
the halo.
These conjectures need to be tested on 
other, more realistic or better dynamically founded models 
(e.g., those of Evans 1994)  of the halo than 
the ones discussed in the present paper.


\subsection*{Acknowledgements}
The author thanks Jesper Sommer-Larsen of TAC for numerous
enlightening discussions and gratefully acknowledges hospitality
of the Aspen Center for Physics, where the earliest version of this
paper was conceived.
This research was generously supported by Danmarks Grundforskningsfond
through its establishment of the Theoretical Astrophysics Center.

\appendix
\section*{Appendix: Position and motion of the Earth 
 relative to the Galaxy and the Large Magellanic Cloud}

This appendix contains formulae (with derivations) giving
positions and proper motions in the rest frame of the Galaxy
of points along the Earth-Large Magellanic Cloud line of sight.

We denote by $\hat{\bf{x}}$, $\hat{\bf{y}}$ and $\hat{\bf{z}}$
unit vectors directed, respectively, toward the Galactic centre (GC), 
tangentially along the rotation of the `local standard of rest'
(LSR) around the galactic centre, and toward the north Galactic pole.
If the velocity of a source relative to the Sun is given
in terms of Galactic-coordinate components  $v_{\rm rad}$, $v_l$ and
$v_b$, the conversion to the $x$, $y$, $z$ components is
\begin{eqnarray}
\label{vel.rel.sun}
{\bf v}'_{\rm s} &=&\hspace{0.3cm}
         (\cos b \cos l \, v_{\rm rad} -\sin b\cos l\, v_b
 -\sin l \, v_l )\: \hat{\bf x}
  \nonumber \\
 & & +\, (\cos b \sin l \, v_{\rm rad} -\sin b\sin l\, v_b
 +\cos l \, v_l )\: \hat{\bf y}
  \nonumber \\
 & & +\, (\sin b\, v_{\rm rad}  + \cos b\, v_b )\: \hat{\bf z}
\end{eqnarray} 
in the rest frame of the Sun.
From the  1990 data of  B. Jones 
(as quoted in Greist 1991) 
the proper motion of the LMC is
\begin{eqnarray}
\label{proper.LMC}
v_{\rm rad} &=&  250\pm 5\,{\rm km}/{\rm s} \nonumber \\
v_b &=&  335\pm 62\,{\rm km}/{\rm s} \nonumber \\
v_l &=&  -31\pm 62\,{\rm km}/{\rm s}.
\end{eqnarray}
Disregarding the error bars and using $D=55\,$kpc, $b=-32^{o}.8$, $l=281^{o}$,
[thus ${\bf r}_{\rm LMC} = D\hat{\bf k} = 55\,{\rm kpc}\,
(0.160,-0.825,-0.542)$; $\hat{\bf k}\cdot\hat{\bf k}=1$]
we arrive at ${\bf v}'_{\rm LMC} = (44,-390,146)\,$km/s.  
Bearing in mind that the Sun's velocity relative to the Galaxy is
${\bf v}_{\odot} = (0,220,0) + (9,11,16) = (9,231,16)\,$km/s, 
where the first and second 
terms are the velocity of LSR relative to the Galaxy and
that of the Sun relative to LSR, the velocity of the LMC
relative to the Galaxy is 
\begin{equation}
\label{vel.LMC}
{\bf v}_{\rm LMC} = {\bf v}_{\odot} + {\bf v}'_{\rm LMC} =
(53,-159,162)\, {\rm km}/{\rm s}.
\end{equation}
[After completing the calculations described in this paper,
 the author learned about more recent values of Jones, Klemola
 \& Lin (1994) who give  ${\bf v}_{\rm LMC} = (60\pm 59, -155\pm 25,
 144\pm 51)\,$km/s.  On the other hand, Kroupa \& Bastian (1997)
 give an estimate based, among other sources, on the Hipparcos
 data: ${\bf v}_{\rm LMC} = (41\pm 44, -200\pm 31, 169\pm 37)\,$km/s.
 These values do not differ sufficiently from velocity (\ref{vel.LMC})
 to lead to a significant change in the results reported in
 this paper.]

Along the Sun-LMC line of sight we can introduce 
auxiliary coordinates defined by the following
orthonormal system: as above, $\hat{\bf k}$ is directed along the 
line of sight, $\hat{\bf\imath}$ lies in the Sun-GC-LMC
plane and poins in the direction of the GC and $\hat{\bf\jmath}= 
\hat{\bf k}\times\hat{\bf\imath}$.  Given the data listed above,
$\hat{\bf\imath} = (0.987, 0.134, 0.088)$ and
$\hat{\bf\jmath} = (0, -0.549, 0.836)$.
Each point on the line of sight
moves relative to the galaxy with velocity ${\bf s}(z) = {\bf v}_{\odot}
(1 -z) + {\bf v}_{\rm LMC} z$ whose components along $\hat{\bf\imath}$
and $\hat{\bf\jmath}$ are
\begin{eqnarray}
\label{sij}
s_i &=& v_{{\odot}_i} (1-z) + v_{{\rm LMC}_i} z \nonumber \\
s_j &=& v_{{\odot}_j} (1-z) + v_{{\rm LMC}_j} z,
\end{eqnarray}
where  $v_{{\odot}_i} =41\,$km/s, $v_{{\rm LMC}_i}=45\,$km/s,
$v_{{\odot}_j}=-113\,$km/s and $v_{{\rm LMC}_j}=223\,$km/s. 

In the above we have neglected the motion (revolution) 
of the Earth (detector `1')
and the satellite (detector `2') around the Sun:
one should in principle 
add the term ${\bf\omega}\times {\bf r}_{\oplus}$
(or the corresponding term for the satellite) to ${\bf v}_{\odot}$.
To avoid (at this stage) unnecessary complication we will keep
disregarding the revolution around the Sun.  For specific
computations in this paper we assume that the `1'-`2' line  points along
the unit vector $\hat{\bf r}$ directed along
$\hat{\bf\omega}\times \hat{\bf k}$.
Since in galactic coordinates $\hat{\bf\omega} = (-0.095, 0.862, 0.498)$,
(derived from data in Mihalas \& Binney 1981)
we find $\hat{\bf r} = (-0.652, 0.324, -0.685)$ and thus
$\hat{\bf r}\cdot\hat{\bf\imath} = -0.660 =\cos\phi_o$ and
$\hat{\bf r}\cdot\hat{\bf\jmath} = -0.750 =\sin\phi_o$ corresponding
to  $\phi_o = 229^o$.

\end{document}